\documentclass{aa}
\usepackage{graphicx}
\usepackage{psfig}
\usepackage{amssymb}
\usepackage{txfonts}

\def\cm3{cm$^{-3}$}

\def\tz{T$_{\rm Z}$}
\def\ngc7027{\object{NGC\,7027}}
\def\ngc7662{\object{NGC\,7662}}
\def\n6741{\object{NGC\,6741}}
\def\n2440{\object{NGC\,2440}}
\def\n5315{\object{NGC\,5315}}
\def\n6445{\object{NGC\,6445}}
\def\n6537{\object{NGC\,6537}}
\def\n6302{\object{NGC\,6302}}
\def\he2{\object{He\,2-11}}
\def\n6543{\object{NGC\,6543}}

\begin{document}


   \title{Probing AGB nucleosynthesis via accurate Planetary Nebula 
        abundances}
 
   \author{P. Marigo\inst{1}
        \and J. Bernard-Salas\inst{2,3} 
        \and S. R. Pottasch\inst{3} 
        \and A. G. G. M. Tielens\inst{2,3}
        \and P. R. Wesselius\inst{2,3}}

   \institute{Dipartimento di Astronomia, Universit\`a di Padova, 
                Vicolo dell'Osservatorio 2, 35122 Padova, Italy
        \and SRON National Institute for Space Research, P.O. Box 800, 
        NL 9700 AV Groningen, The Netherlands
        \and  Kapteyn Astronomical Institute, P.O. Box 800, 9700 AV 
        Groningen, The Netherlands }

   \titlerunning{Probing AGB nucleosynthesis via PN abundances}
   
   \offprints{P. Marigo \\ \email{marigo@pd.astro.it}}

  \date{Received ; accepted}

   \abstract{
The elemental abundances of ten planetary nebulae, derived with high accuracy 
including ISO and IUE spectra, are analysed with the aid of 
synthetic evolutionary models
for the TP-AGB phase.
The accuracy on the observed abundances is essential in order to make
a reliable comparison with the models. The advantages of the
infrared spectra in achieving this accuracy are discussed.
Model prescriptions are varied until we  achieve the 
simultaneous reproduction of all elemental features, which allows  
placing important constraints on the characteristic 
masses and nucleosynthetic processes experienced by the stellar progenitors.
First of all, it is possible to separate the sample into two groups of PNe, 
one indicating the occurrence of only the third dredge-up during the 
TP-AGB phase, and the other showing also the chemical signature of hot-bottom 
burning.
The former group is reproduced by stellar models 
with variable molecular opacities (see Marigo 2002), adopting initial solar
metallicity, and typical efficiency of the third dredge-up, 
$\lambda \sim 0.3-0.4$. 
The latter group of PNe, with extremely high He content 
($0.15 \le$He/H$\le 0.20$) and marked oxygen deficiency, 
is consistent with original sub-solar metallicity
(i.e. LMC composition). Moreover, we are able to explain quantitatively both 
the N/H--He/H correlation and  the N/H--C/H  anti-correlation, 
thus solving the discrepancy pointed out long ago by Becker \& Iben (1980).
This is obtained only under the hypothesis that 
intermediate-mass TP-AGB progenitors ($M \ga 4.5-5.0\, M_{\odot}$) with LMC 
composition have suffered a number of very efficient, carbon-poor, 
dredge-up events.
Finally, the neon abundances of the He-rich PNe can be recovered 
by invoking a significant production of $^{22}$Ne during thermal pulses,
which would imply a reduced role of the $^{22}$Ne($\alpha$, n)$^{25}$Mg 
reaction as neutron source to the s-process nucleosynthesis in these stars.
%
   }
        
   \maketitle
%

\section{Introduction}
\label{sec_intro}
Planetary Nebulae (PNe) are assumed to consist of
the gas ejected via stellar winds by low-
and intermediate-mass stars (having initial masses $0.9 \le M/M_{\odot} 
\le M_{\rm up}$, with  $M_{\rm up} \sim 5-8 M_{\odot}$ 
depending on model details) during their last evolutionary stages,
the so-called Thermally Pulsing Asymptotic Giant Branch (TP-AGB) phase.

PNe offer potentially a good possibility to test the results of 
stellar nucleosynthesis. This can be done in a reliable way by comparing
the predicted abundances of the gas ejected close to the end 
of the AGB phase with the observed abundances 
because the expelled hot gas
remains unaffected by interaction with the ISM or with previous
shell ejection. Furthermore the ionized gas surrounding the central star 
shows lines of many elements from which accurate abundances can
be derived. Also by the ejection of the outer layers PNe contribute
to the enrichment of the interstellar medium (ISM) and therefore, a
knowledge of these processes are essential to better understand the
chemical composition of the Galaxy.

PN elemental abundances represent the cumulative record of all nucleosynthetic
and mixing processes that may have changed the original composition of the
gas since the epoch of stellar formation. 
In fact, stellar evolution models predict the occurrence 
of several episodes 
in which the envelope chemical composition is altered by mixing with nuclear
products synthesised in inner regions and brought up to the surface by 
convective motions (e.g. Iben \& Renzini 1983; Forestini \& Charbonnel 1997;
Girardi et al. 2000).
These dredge-up events usually take place when a star reaches its Hayashi
line and develops an extended convective envelope, either during the 
ascent on the Red Giant Branch (RGB; the first dredge-up), 
or later on the early AGB (the second dredge-up).
Then, the subsequent TP-AGB evolution is characterised by a rich 
nucleosynthesis whose products may be recurrently exposed to the surface
synchronised with thermal pulses  
(the third dredge-up), or convected upward
from the deepest envelope layers of the most massive stars 
(hot-bottom burning, hereinafter also HBB).

As a consequence, the surface abundances of several elements 
(e.g. H, He, C, N, O, Ne, Mg) may be significantly altered, to an extent
that crucially depends on stellar parameters (i.e. mass and metallicity)
various incompletely understood physical 
processes (e.g. convection, mass loss), and model input prescriptions 
(e.g. nuclear reaction rates, opacities, etc.).
In this sense, the interpretation
of the elemental patterns observed in PNe should give a good insight 
into the evolutionary and nucleosynthetic properties of the stellar 
progenitors, thus putting constraints on these processes.

For instance, the enrichment in C exhibited by some PNe should give 
a measure of the efficiency of the third dredge-up, still a 
matter of debate.
Conversely, the deficiency in C shown by some PNe with 
N overabundance could
be interpreted as the imprint of HBB, implying rather massive TP-AGB 
stellar progenitors (with initial masses $\ge 4\, M_{\odot}$).    
The He content should measure the cumulative effect 
produced by the first, possibly second and third dredge-up, and HBB.
The O abundance may help to constrain the chemical composition
of the convective inter-shell developed at thermal pulses,
as well as the efficiency of HBB, depending on
whether this element is found to be preserved, enhanced or depleted 
in the nebular composition. 
Finally, the Ne abundance in PNe could give important information
about the synthesis of this element during thermal pulses, 
i.e. providing an indirect estimate of the efficiency of the  
$^{22}$Ne($\alpha$, n)$^{25}$Mg reaction 
with important implications for the slow-neutron capture nucleosynthesis.

In this context, the present study aims at investigating the above
issues by analysing accurate determinations of elemental abundances
for a sample of PNe with the aid of stellar models for low-
and intermediate stars, that follow their evolution from the main sequence
up to the stage of PN ejection.
Particular attention is paid to modelling the TP-AGB phase to
derive indications on the mass range 
and the metallicity of the stellar progenitors involved, and 
the related  nucleosynthetic processes, i.e. the second and 
third dredge-up and HBB. 

An important problem when deriving PNe abundances is the
correction for unseen stages of ionization. When using optical
and/or UV spectra many stages of ionization are missing and have to
be inferred, making the error in the abundance determination high.
This estimate for unobserved stages of ionization is done making use
of Ionization Correction Factors (ICF) which are mainly derived on
the basis of similarities of ionization potentials or ionization
models. The latter need a very good knowledge of the stellar
parameters which is often not known. When all important stages of
ionization of a certain element are measured no ICF is needed (or
ICF=1). In the past literature ICF ranging from 2-5 (and sometimes
20) are found. Many missing stages of ionization are seen in the
infrared providing an important complement to the UV and optical
spectra. The ICF of many elements has been drastically reduced
thanks to the inclusion of the ISO (Kessler et al. 1996) data. It has certainly improved
the Ne, Ar, Cl and S abundances and has provided information of
other important stages of ionization such as C$^{++}$, O$^{3+}$ and
N$^{++}$.  In many cases the ICF is not needed and on the others is
often lower than 1.5. Another important advantage is the
independence of the infrared lines to the adopted electron
temperature.  This avoids uncertainties when the electron
temperature adopted to derive the abundances is uncertain or when
there are electron temperature fluctuations in the nebula. These and
other advantages have been previously discussed by Beintema \&
Pottasch (1999) and Bernard Salas et al. (2001).

The paper is organised as follows.
Section~\ref{sec_obs} introduces the sample of ten planetary nebulae,
in terms of individual characteristic like:  galactic coordinates, 
radii, nebular fluxes in H and \ion{He}{ii} recombination lines,
Zanstra temperatures and luminosities of the central nuclei.
These two latter parameters locate the central stars 
in the Hertzsprung-Russell (HR) diagram. 
Section~\ref{sec_obsab} presents  the nebular elemental abundances 
of He, C, N, O, Ne, S, and Ar, compared with the solar values, 
and subgrouped as a function of  helium content.
Section~\ref{sec_dups} outlines a summary of the main physical processes 
expected to alter the surface 
chemical composition of low- and intermediate-mass stars.
Section~\ref{sec_tpagb}  details the synthetic TP-AGB models adopted 
for our theoretical study, in terms of the main  
input parameters. The interpretative analysis of the abundance data 
is developed in Sect.~\ref{sec_obsmod}. 
Finally, a recapitulation of the most relevant
conclusions and implications in Sect.~\ref{sec_concl} closes the paper.    

\section{Sample of PNe}
\label{sec_obs}
\subsection{General}
  The sample used is biased to bright objects, in order to measure
  many different stages of ionization and accurately
  derive their abundances.  The general parameters of the PNe used in
  this study are given in Table\,\ref{general}.  References for the
  assumed distances, magnitudes  and extinction are given as
  footnotes in the table.\\
  
\begin{table*}
  \caption{Parameters of the PNe. The M$_V$ is not corrected for
    extinction. References to distance, V magnitude and extinction
    are given by superscripts {\em dx, mx,} and {\em ex} respectively.}
  \label{general}
  \begin{center}
    \begin{tabular}{l c c c c c  }
      \hline
      \hline
      \makebox{\rule[0cm]{0cm}{0.33cm}Name} & l(\degr) & b(\degr) & d
      (kpc) & M$_V$ (mag)  &
      E$_{B-V}$ \\
      \hline
      
NGC\,2440$^{d1,m1,e1}$ & 234.8 &  2.42 & 1.63 & 17.49 & 0.34  \\     
NGC\,5315$^{d2,m2,e2}$ & 309.1 & -4.40 & 2.60 & 14.40 & 0.37  \\
NGC\,6302$^{d3,m3,e3}$ & 349.5 &  1.06 & 1.60 & 18.90:& 0.88  \\
NGC\,6445$^{d1,e4,e4}$ & 8.08  &  3.90 & 2.25 & 19.04 & 0.72  \\
NGC\,6537$^{d1,m4,e5}$ & 10.10 &  0.74 & 1.95 & 22.40 & 1.22  \\ 
NGC\,6543$^{d4,m1,e6}$ & 96.47 &  29.9 & 1.00 & 11.29 & 0.07  \\
NGC\,6741$^{d1,m5,e7}$ & 34.6  & -2.28 & 1.65 & 19.26 & 0.75  \\
NGC\,7027$^{d5,m1,e8}$ & 84.93 & -3.50 & 0.65 & 16.53 & 0.85  \\
NGC\,7662$^{d3,m1,e7}$ & 106.6 & -17.6 & 0.96$^\ast$ & 14.00 & 0.12  \\
He\,2-111$^{d1,m3,e5}$ & 315.0 & -0.37 & 2.50 & 20.00:& 0.77  \\

      \hline
    \end{tabular}
  \end{center}
References: $^{d1}$ Average from Acker et al. (1992); $^{d2}$ Liu et
al. (2001); $^{d3}$ Terzian
(1997);
 $^{d4}$ Reed et al. (1999); $^{d5}$ Bains et al. (2003);
$^{m1}$ Ciardullo et al. (1999); $^{m2}$ Acker et al. (1992); $^{m3}$ Assumed M$_V$;
$^{m4}$ Pottasch (2000);
$^{m5}$ Heap et al. (1989); $^{e1}$ Bernard Salas et al. (2002);
$^{e2}$ Pottasch et al. (2002);
$^{e3}$ Beintema \& Pottasch (1999); $^{e4}$ van Hoof et al. (2000);
$^{e5}$ Pottasch et al. (2000);
$^{e6}$ Bernard Salas (2003); $^{e7}$ Pottasch et al. (2001); $^{e8}$ Bernard
Salas et al. (2001).\\
: Large error.\\
$^\ast$ Average distance by Terzian (1997)


\end{table*}

  \begin{figure*}[]
  \begin{center}
    \begin{picture}(500,250)(-80,15)
      \psfig{figure=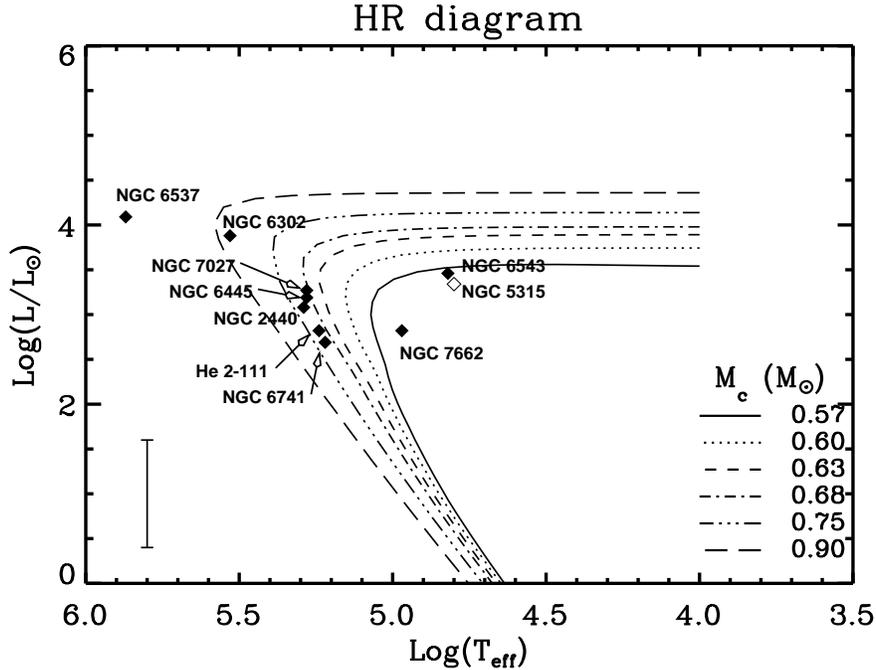,width=350pt,angle=-90}
    \end{picture}
  \end{center}
  \caption{HR diagram for the PNe of the sample (diamonds). The
    $T_{\mathrm{eff}}$~have been derived with the Zanstra method using
    the helium lines except for \object{NGC\,5315} (open diamond).
    The Post-AGB evolutionary tracks from Vassiliadis \& Wood
    (1994) for Z=0.016 are also plotted for different core masses, 
    indicated in the
    lower-right corner of the figure. In the lower-left the
    uncertainty in the luminosity due to the error of a factor 
    two in the distance is shown}
  \label{hr}
\end{figure*}

\begin{table*}
    \caption{Radius, Zanstra temperature and luminosity of the observed
      PNe. F(H$\beta$) and F(4684\AA) are in units of 10$^{-12}$ erg~cm$^{-2}$~s$^{-1}$ (not corrected
      for extinction). The radius is in meter.}
   \label{tz}
   \begin{center}
    \begin{tabular}{l|cccc|  cccc}
      \hline
      \hline
      Name & \multicolumn{4}{c|}{Hydrogen} & 
      \multicolumn{4}{c}{Helium}\\ \cline{2-5} \cline{6-9}
      & \makebox{\rule[0cm]{0cm}{0.43cm}F(H$\beta$)}$^\dagger$ & Radius & $\log (\frac{L_*}{L_{\odot}})$ &$\log T_Z(H)$
       & F(4686\AA)$^\ast$& Radius &  $\log (\frac{L_*}{L_{\odot}})$ &$\log T_Z(He)$\\
      \hline

NGC\,2440 & 31.6 & 2.2E+07 & 2.95 & 5.25 &  19.3 & 2.1E+07 & 3.08 & 5.29 \\
NGC\,5315 & 38.1 & 2.7E+08 & 3.34 & 4.80 &  -    &      -  &    - & -    \\
NGC\,6302:& 29.5 & 1.7E+07 & 3.86 & 5.52 &  16.6 & 1.7E+07 & 3.88 & 5.53 \\
NGC\,6445 & 7.60 & 2.4E+07 & 3.19 & 5.28 &  3.40 & 2.4E+07 & 3.19 & 5.28 \\
NGC\,6537 & 2.20 & 5.8E+06 & 3.50 & 5.67 &  3.10 & 4.6E+06 & 4.09 & 5.87 \\
NGC\,6543 & 245  & 3.6E+08 & 2.96 & 4.64 &  14.7 & 2.8E+08 & 3.46 & 4.82 \\
NGC\,6741 & 4.40 & 1.8E+07 & 2.69 & 5.22 &  1.60 & 1.8E+07 & 2.69 & 5.22 \\
NGC\,7027 & 75.9 & 2.7E+07 & 3.31 & 5.29 &  31.1 & 2.8E+07 & 3.27 & 5.28 \\
NGC\,7662 & 102  & 7.8E+07 & 2.54 & 4.87 &  17.4 & 6.8E+07 & 2.82 & 4.97 \\
He\,2-111:& 0.98 & 2.5E+07 & 2.37 & 5.08 &  0.89 & 2.0E+07 & 2.82 & 5.24 \\

      \hline
    \end{tabular}
   \end{center}
: Large error in the radius, temperature and luminosity.\\
$^\dagger$ From the same references as the  extinction in Table\,\ref{general}.\\ 
$^\ast$ From Acker et al. (1992).

\end{table*}

  Galactic coordinates show that most of these nebulae belong to the
  disk (except \object{NGC\,6543} and \object{NGC\,7662}) and could be
  descendants of young progenitors.  Distances are very uncertain and
  great care was taken to adopt the most reliable ones from the
  literature. They vary between 0.8 and 2.5 kpc and therefore are
  close, as would be expected for bright objects. There are different
  values in the literature that do not agree within the uncertainties
  the authors quote.  Note that the M$_V$ of \object{He\,2-111} and
  \object{NGC\,6302} are assumed since their central stars have never
  been seen.  The extinction is low in most cases except for
  \object{NGC\,6537}. In order to classify them according to their
  chemical composition we can investigate the C/O ratio together with
  the He/H abundance in Table\,\ref{abun}.  
  There are two C-rich PNe, six O-rich PNe and
  two for which it is difficult to assess their nature since the C/O
  ratio is (although lower than one) very close to unity (within the
  uncertainties).  Notice that this sample contains a higher percentage of 
  PNe with a high He/H ratio than many other samples.
     
  \subsection{HR diagram}
\label{ssec_hrd}

 With the data in Table\,\ref{general} and the H$\beta$ and helium
  $\lambda$4686\AA~fluxes the Zanstra temperatures (\tz), radii and
  luminosities have been derived (see Table\,\ref{tz}). As pointed out
  by Stasi\'nska \& Tylenda (1986) when using the Zanstra method, 
  \tz~is over-estimated in the case of hydrogen and underestimated when
  using helium. This is because the Zanstra method assumes that
  energies above 54.4 eV are only absorbed by helium. This is not
  completely true. In addition recombination of He$^{2+}$ sometimes produces
  more than one photon which can ionize hydrogen and the proportion of
  stellar photons with energies above 54.4 increases with T$_{eff}$.
  Both \tz(He) and \tz(H) yield the same results for most PNe. \tz(H)
  fails when the nebula is thin and some photons escape. In the case
  of a thick nebula both methods should yield the same result, but
  this is tricky because a nebula can be thick in the torus and thin
  in the poles.
  For all those reasons the \tz(He) was preferred over \tz(H).
  
  These results are shown in Fig.\,\ref{hr}. For \object{NGC\,5315} no
  helium line is detected so that results using \tz(H) have been
  plotted.  The evolutionary tracks of Vassiliadis \& Wood (1994) are
  related to the core mass rather than the initial mass. Stars with
  different initial mass
  and different mass loss functions can lead to the same core mass.

  The majority of PNe are within the range of temperature and
  luminosity of the theoretical evolutionary tracks. Most are in the
  last stage of the PNe phase.  \object{NGC\,6537} lies outside and it
  must be noticed that using  \tz(H) makes this object 
  closer to \object{NGC\,6302}. While this figure may provide some
  insight in core mass and time evolution of these objects, the many
  uncertainties (especially in distance) should be born in mind.
 
\section{Accurate abundances of ISO-observed PNe}

  \subsection{Abundances}
  \label{sec_obsab}  
  Accurate abundances are needed in order to make a reliable
  comparison with theoretical models. For this purpose a sample of ten
  PNe was selected in which precise abundances using ISO data have been
  derived. The references for the abundances are as follows: NGC\,2440
  $\rightarrow$ Bernard Salas et al. (2002); NGC\,5315 $\rightarrow$
  Pottasch et al.  (2002); NGC\,6302 $\rightarrow$ Pottasch \&
  Beintema (1999); NGC\,6445 $\rightarrow$ van Hoof et al. (2000);
  NGC\,6537 and He\,2-111 $\rightarrow$ Pottasch et al. (2000);
  NGC\,6543 $\rightarrow$ Bernard-Salas et al. (2003); NGC\,6741 and
  NGC\,7662 $\rightarrow$ Pottasch et al. (2001); NGC\,7027
  $\rightarrow$ Bernard Salas et al. (2001) and Bernard Salas et
    al. (in prep).
  
\begin{table*}[]
  \begin{center}
  \caption{PNe abundances$^\natural$ w.r.t. hydrogen in number. 
The number between parenthesis (x) stands for 10$^x$. 
Abundance reference values for the Sun, Orion, and LMC are also included.}
  \label{abun}
  \begin{tabular}{lccccccc}
    \hline
    \hline
    Name & Helium & Carbon(-4) & Nitrogen(-4) & Oxygen(-4) & Neon(-4) &
    Sulfur(-5) & Argon(-6)\\ 
    \hline

\object{NGC\,2440}&  0.119 & 7.2  & 4.4  & 3.8  &  1.1  &  0.5  &  3.2\\
\object{NGC\,5315}&  0.124 & 4.4  & 4.6  & 5.2  &  1.6  &  1.2  &  4.6\\
\object{NGC\,6302}&  0.170 & 0.6  & 2.9  & 2.3  &  2.2  &  0.8  &  6.0\\
\object{NGC\,6445}&  0.140 & 6.0  & 2.4  & 7.4  &  2.0  &  0.8  &  3.8\\
\object{NGC\,6537}&  0.149 & 1.8  & 4.5  & 1.8  &  1.7  &  1.1  &  4.1\\  
\object{NGC\,6543}&  0.118 & 2.5  & 2.3  & 5.5  &  1.9  &  1.3  &  4.2 \\
\object{NGC\,6741}&  0.111 & 6.4  & 2.8  & 6.6  &  1.8  &  1.1  &  4.9\\
\object{NGC\,7027}&  0.106 & 5.2  & 1.5  & 4.1  &  1.0  &  0.9  &  2.3\\
\object{NGC\,7662}&  0.088 & 3.6  & 0.7  & 4.2  &  0.6  &  0.7  &  2.1\\ 
\object{He\,2-111}&  0.185 & 1.1  & 3.0  & 2.7  &  1.6  &  1.5  &  5.5\\
Sun$^\ast$        &  0.100 & 3.55 & 0.93 & 4.9  &  1.2  &  1.86 &  3.6\\
Orion$^\dagger$   &  0.098 & 2.5  & 0.60 & 4.3  &  0.78 &  1.5  &  6.3\\
$<{\rm LMC}>^\sharp$    &  0.089 & 1.10 & 0.14 & 2.24 &  0.41 &  0.65 &  1.9\\ 
    \hline
  \end{tabular}
\end{center}
$^\natural$ See Sec.\,\ref{sec_obsab} for references.\\
$^\ast$ Grevesse \& Sauval (1998) and Anders \& Grevesse (1989) except
the oxygen abundance which was taken  from
Allende Prieto et al. (2001).\\
$^\dagger$  Esteban et al. (1998).\\  
$^\sharp$ From Dopita et al. (1997).\\
\end{table*}

\begin{figure*}[]
  \begin{center}
    \begin{picture}(450,250)(0,0)
      \psfig{figure=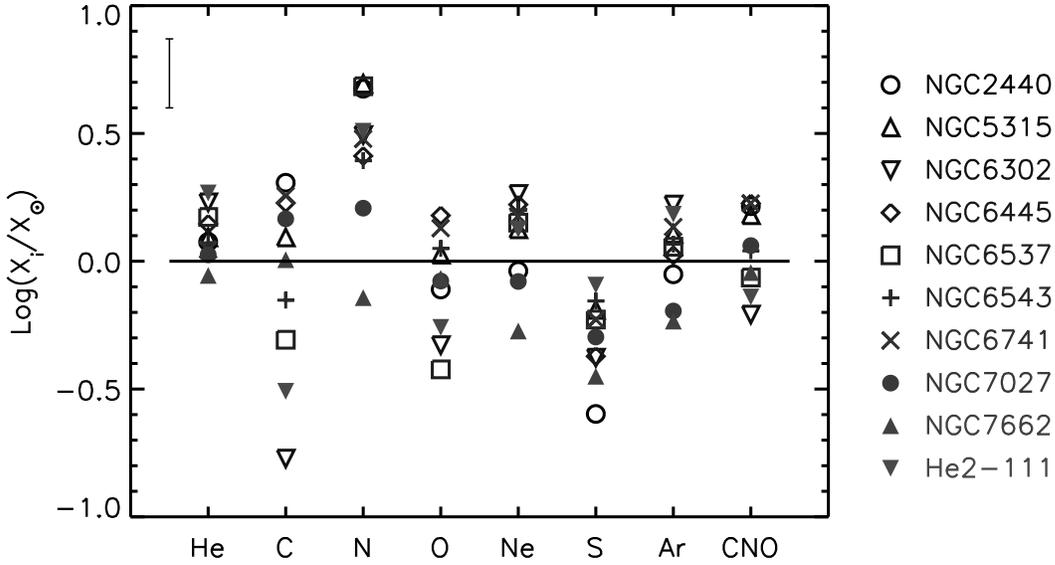,width=450pt}
    \end{picture}
  \end{center}
  \caption{Observational abundances with respect  to Solar. In the
    upper-left corner a typical error bar for all elements except helium is plotted. The solid line
    represents abundances equal to solar abundances}
  \label{solar}
\end{figure*}

  It should be noticed that  NGC\,6302, NGC\,6537 and He\,2-111
    are among those PNe with  a hot central star, a strong
  bipolar morphology, and in which high velocity shocks are
  present. The presence of the latter can affect the abundance
  composition which has been derived assuming that the ionization is
  produced by the hot central
  star. For He\,2-111 the velocity of the shocks have been estimated
  by Meaburn \& Walsh (1989) to be $\sim$380 km/s. They show that these high
  velocity shocks are
  localized in the outermost parts of the bi-polar lobes (which are
  less dense). They  conclude that photoionization by a hot star is
  the most plausible dominant process in the core and dense
  disk. Since  these shocks affect the less denser
  regions the effect that they might have on the abundance
  determination is very small and can be neglected.
  NGC\,6302 has velocity shocks of $\sim$500 km/s (Lame \& Ferland
  1991). They fit photoionization models to the spectrum which 
  indicate that only very high stages of ionization, such as Si${^6+}$
  (IP 167 eV), are produced by shocks. These high stages of ionization
  have negligible weight on the overall abundance which  is  therefore not
   affected by the presence of the shocks. The speed of
   the shocks in NGC\,6537 is $\sim$ 300 km/s (Corradi \& Schwarz
   1993). Hyung (1999) studied this specific problem of the source of
   the nebular emission arising either by the hot central star or by
   shock heating, with the
   help of photoionization models. He concludes that  the radiation
   of the hot central star is responsible for the emission. In
   summary, the high velocity shocks  present in these PNe do not
   play a role in the abundance determination.

  The error in the abundances in Table\,\ref{abun} is about 20 to
  30\%.  These errors only include the uncertainty in the intensities
  of the lines used to derive the abundances. Other errors, e.g.
  using Ionization Correction Factor (ICF) to derive abundances or the
  effect of uncertainties in the atomic parameters (especially the
  collisional strengths) are difficult to quantify. The abundances
  shown in Table\,\ref{abun} have been derived using an ICF near
  unity. Often all important stages of ionization are observed, in
  which case no error from the ICF is involved. The abundances have been
  obtained using the most recent results available for the collisional
  strengths (mainly from the IRON project) so that these uncertainties
  are much reduced.  A quantitative idea of the errors on the
  abundance can also be inferred by looking at the sulfur and argon
  abundances in Fig.\,\ref{solar}.  Sulfur and argon are not supposed
  to vary in the course of evolution, and as can be seen from the
  figure, they are within 30\%. This strongly suggests that the
  abundance error derived from the intensity of the lines is the main
  contributor to the total error.  Therefore 30\% is a good
  estimate of the abundance error and has been assumed in this work
  for all nebulae except for the nitrogen abundance of NGC\,6543 which
  has an error 50\%. This is because the main contribution to the
  nitrogen abundance comes from the N$^{++}$ whose  ionic abundance is
  derived either with the 57.3 $\mu$m or 1750 \AA~lines. The former is
  density dependent and the second temperature dependent, increasing
  therefore the error. The error in the helium abundance is around 5\%
  except, again, for NGC\,6543 where it is 7-8\%.

  \subsection{Comparison with Solar abundances}
  \label{solartext}
  
  The PNe abundances (Table \ref{abun}) are shown in Fig.\,\ref{solar}
  with respect to the solar. The typical error bar applies to all
  elements except helium for which the error is 6 times smaller. 
  This comparison is
  important because in principle one might expect that the progenitor
  stars of these PNe have evolved from a solar metallicity.
  Therefore, primary elements that do not change in the course of
  evolution should lie close to the solar line, elements that are
  destroyed or produced should
  lie below or above.

 An inspection of the figure leads to several conclusions.
  \object{NGC\,7662} shows low abundances for all elements.  It is
  then suspected that the progenitor mass of this PNe was low and that
  it
  did not experience much change in the course of evolution.\\
  
  All other PNe show a He enhancement, which should be expected as He
  is brought to the surface in the different dredge-up episodes.  Four
  PNe show a decrease in carbon, especially \object{NGC\,6302} and
  \object{He\,2-111}. The remaining PNe show solar or enhanced carbon.
  All PNe (except \object{NGC\,7662}) show an increase of nitrogen.
  The oxygen abundance is close to solar for all PNe except three.
  The exceptions are \object{NGC\,6302}, \object{NGC\,6537} and
  \object{He\,2-111} which show a clear decrease.  It should be
  noticed that the solar abundance adopted in this work is that of
  Allende Prieto et al. (2001). If the value of Grevesse \& Sauval
  (1998) had been adopted all PNe would lie
  below the solar value.
 
  Within the errors all neon abundances except that
  of \object{NGC\,7662} agree with solar. Of all elements neon is
  probably the best determined and it is likely that the error is
  somewhat smaller than 30\%.  In this sense it is interesting to see
  how PNe with higher neon abundances are clumped at around 0.15 and
  the three with lower values also have lower helium abundance
  of the sample.
  
  The remaining elements, sulfur and argon,  are supposed to remain unchanged
  in the course of evolution and therefore should be close to solar.
  The argon abundance is compatible with this but the sulfur abundance
  is lower than solar.

\section{Nucleosynthesis in low- and intermediate-mass stars}
\label{sec_dups}
We will  recall the main nucleosynthetic and convective 
mixing events that may possibly alter the surface chemical composition of a 
low-/intermediate-mass star in the course of its evolution. 
According to the classical scenario\footnote{We do not consider here 
any additional extra-mixing process,
such as the one invoked to explain the observed abundance  anomalies in 
low-mass RGB stars (see e.g. Charbonnel 1995).}
 four processes are important 
(see e.g. Forestini \& Charbonnel 1997 for a recent extended analysis), 
namely:

\paragraph{The first dredge-up.} At the base of the RGB the outer convective 
envelope reaches regions of partial hydrogen burning (CN cycle). 
As a consequence, the surface abundance of $^{4}$He is increased
(and that of H depleted),
$^{14}$N and $^{13}$C are enhanced at the expense of
$^{12}$C, while $^{16}$O remains almost unchanged.

\paragraph{The second dredge-up.} This occurs in stars initially more massive 
than 3-5 $M_{\odot}$ (depending on composition) during the early AGB phase. 
The convective envelope  penetrates into the helium core
(the H-burning shell is extinguished) so that the surface abundances of 
$^{4}$He and $^{14}$N increase, while those of
$^{12}$C, $^{13}$C and $^{16}$O decrease.

\paragraph{The third dredge-up.} This takes place during the TP-AGB evolution 
in stars more massive than $\approx 1.5\, M_{\odot}$ for solar composition, 
starting at lower masses for lower metallicities (see Marigo et al. 1999). 
It actually consists of 
several mixing episodes occurring at thermal pulses during which 
significant amounts of $^{4}$He and $^{12}$C, and smaller 
quantities of other newly-synthesized products 
(e.g. $^{16}$O, $^{22}$Ne, $^{25}$Mg, s-process elements) are convected to
the surface.

\paragraph{Hot bottom burning.} This occurs in
the most massive and luminous 
AGB stars (with initial masses $M \ga 4-4.5 M_{\odot}$, 
depending on metallicity). 
  The convective envelope  penetrates deeply into 
  the hydrogen-burning shell, and the CN-cycle nucleosynthesis actually
  occurs in the deepest envelope layers of the star. 
  As a consequence, besides the
  synthesis of new helium,  $^{12}$C is first converted 
  into  $^{13}$C and then into $^{14}$N. In the case of high
  temperatures and  after a  sufficiently long time, the ON cycle can also  be 
  activated, so that $^{16}$O is burned into $^{14}$N.  

It should be remarked that the third
dredge-up and hot-bottom burning are the processes that are expected to
produce the most significant changes in CNO and He surface abundances, 
being affected at the same time by the largest uncertainties in the
theory of stellar evolution. This latter point motivates the adoption
of free parameters (e.g. the dredge-up efficiency) to describe these processes 
in synthetic TP-AGB models, that are discussed next.

\begin{table*}
\caption{Assumptions adopted in the TP-AGB synthetic calculations 
discussed in the paper.}
\label{tab_mod}
\begin{tabular}{lcclcccc}
\noalign{\smallskip}
\hline
\hline
\noalign{\smallskip}
\multicolumn{1}{c}{} &
\multicolumn{1}{c}{} &
\multicolumn{1}{c}{} &
\multicolumn{1}{c}{$3^{\rm rd}$ D-up efficiency} &
\multicolumn{3}{c}{Inter-shell composition}  & 
\multicolumn{1}{c}{Initial oxygen abundance}\\
\noalign{\smallskip}
\hline
\noalign{\smallskip}
\multicolumn{1}{c}{Models (ref.)} &
\multicolumn{1}{c}{$Z$} &
\multicolumn{1}{c}{Opacity$^1$} &
\multicolumn{1}{c}{$\lambda^2$} &
\multicolumn{1}{c}{$X_{\rm csh}(^{12}$C)} &
\multicolumn{1}{c}{$X_{\rm csh}(^{16}$O)} &
\multicolumn{1}{c}{$X_{\rm csh}(^{4}$He)} &
\multicolumn{1}{c}{$X_0$(O)$^3$}\\
\noalign{\smallskip}
\hline
\noalign{\smallskip}
A) Figs.~\protect{\ref{fig1_z02kfix}}, \protect{\ref{fig_hbbz02}}  
& 0.019 & ${\kappa_{\rm fix}}$ & const. 0.50 & 0.220 & 0.020 & 0.760 & O$_{\odot}$ High\\
B) Fig.~\protect{\ref{fig_che2}}
& 0.019 & ${\kappa_{\rm var}}$ & const. 0.50 & 0.220 & 0.020 & 0.760 & O$_{\odot}$ High\\
C) Fig.~\protect{\ref{fig_che2}}
& 0.019 & ${\kappa_{\rm var}}$ & var. 0.457 & 0.220 & 0.020 & 0.760 & O$_{\odot}$ High\\
D) Fig.~\protect{\ref{fig_che2}}, \protect{\ref{fig1_z02kvar}}, 
\protect{\ref{fig2_z02kvar}}
& 0.019 & ${\kappa_{\rm var}}$ & var. 0.457 & 0.220 & 0.020 & 0.760 & O$_{\odot}$ Low\\
E) Fig.~\protect{\ref{fig_che2}}, \protect{\ref{fig1_z02kvar}}, \protect{\ref{fig2_z02kvar}}
& 0.019 & ${\kappa_{\rm var}}$ & var. 0.30 & 0.220 & 0.020 & 0.760 & O$_{\odot}$ Low\\
F) Figs.~\protect{\ref{fig_hbbz02}}, \protect{\ref{fig1_z02kvar}}, 
\protect{\ref{fig2_z02kvar}}
& 0.019 & ${\kappa_{\rm var}}$ & var. 0.88, 0.96 & 0.050 & 0.005 & 0.945 & O$_{\odot}$ Low\\
G) Fig.~\protect{\ref{fig_hbbz008}} 
& 0.008 & ${\kappa_{\rm fix}}$ & const. 0.50 & 0.220 & 0.020 & 0.760 & $(Z/Z_{\odot}) \times$ (O$_{\odot}$ High) \\
H) Fig.~\protect{\ref{fig_hbbz008}} 
& 0.008 & ${\kappa_{\rm fix}}$ & const. 0.90 & 0.220 & 0.020 & 0.760 & $(Z/Z_{\odot}) \times$ (O$_{\odot}$ High) \\
I) Fig.~\protect{\ref{fig_hbbz008}} 
& 0.008 & ${\kappa_{\rm var}}$ & const. 0.90 & 0.220 & 0.020 & 0.760 & $(Z/Z_{\odot}) \times$ (O$_{\odot}$ High) \\
J) Figs.~\protect{\ref{fig_hbbz008}}, 
\protect{\ref{fig1_z02kvar}}, \protect{\ref{fig2_z02kvar}}
& 0.008 & ${\kappa_{\rm var}}$ & const. 0.90 & 0.030 & 0.001 & 0.969 &  $(Z/Z_{\odot}) \times$ (O$_{\odot}$ High)\\
K) Figs.~\protect{\ref{fig_hbbz008}}, \protect{\ref{fig1_z02kvar}}, 
\protect{\ref{fig2_z02kvar}} 
& 0.008 & ${\kappa_{\rm var}}$ & const. 0.90 & 0.010 & 0.001 & 0.999 &  $(Z/Z_{\odot}) \times$ (O$_{\odot}$ High) \\
\noalign{\smallskip}
\hline
\noalign{\smallskip}
\end{tabular}
\begin{minipage}{\textwidth}
\normalsize
$^1$ ``$\kappa_{\rm fix}$'' corresponds to solar-scaled molecular opacities
by Alexander \& Ferguson (1994);

~~``$\kappa_{\rm var}$'' denotes the variable molecular opacities 
as calculated according to Marigo (2002).

$^2$ ``const. value'' means that the same constant value of $\lambda$ 
is adopted at each dredge-up episode;

~~``var. value'' means that $\lambda$ is made vary up to
a maximum value, according to the analytic recipe  by Karakas et al. \\
~~(2002, their equation 7)

$^3$ ``O$_{\odot}$ High'' refers the determination of the oxygen abundance 
for the Sun by Anders \& Grevesse (1989); 

~~ ``O$_{\odot}$ Low'' refers to the recent determination 
of the oxygen abundance for the Sun by Allende Prieto et al. (2001).

Other  assumptions common to all TP-AGB models are: \begin{itemize}
\item $\log T_{\rm b}^{\rm dred} = 6.4$ following Marigo et al. (1999);
\item Mixing-length parameter  $\alpha=\Lambda/H_p = 1.68$
(where $H_p$ denotes the pressure-scale height, and $\Lambda$ is the
mixing-length) following
Girardi et al. (2000), unless otherwise specified;
\item Nuclear reaction rates from Caughlan \& Fowler (1988).
\end{itemize}
\end{minipage}

\smallskip
\line(1,0){510}
\end{table*}

\begin{figure*}
\begin{minipage}{0.77\textwidth}
\includegraphics[width=\textwidth]{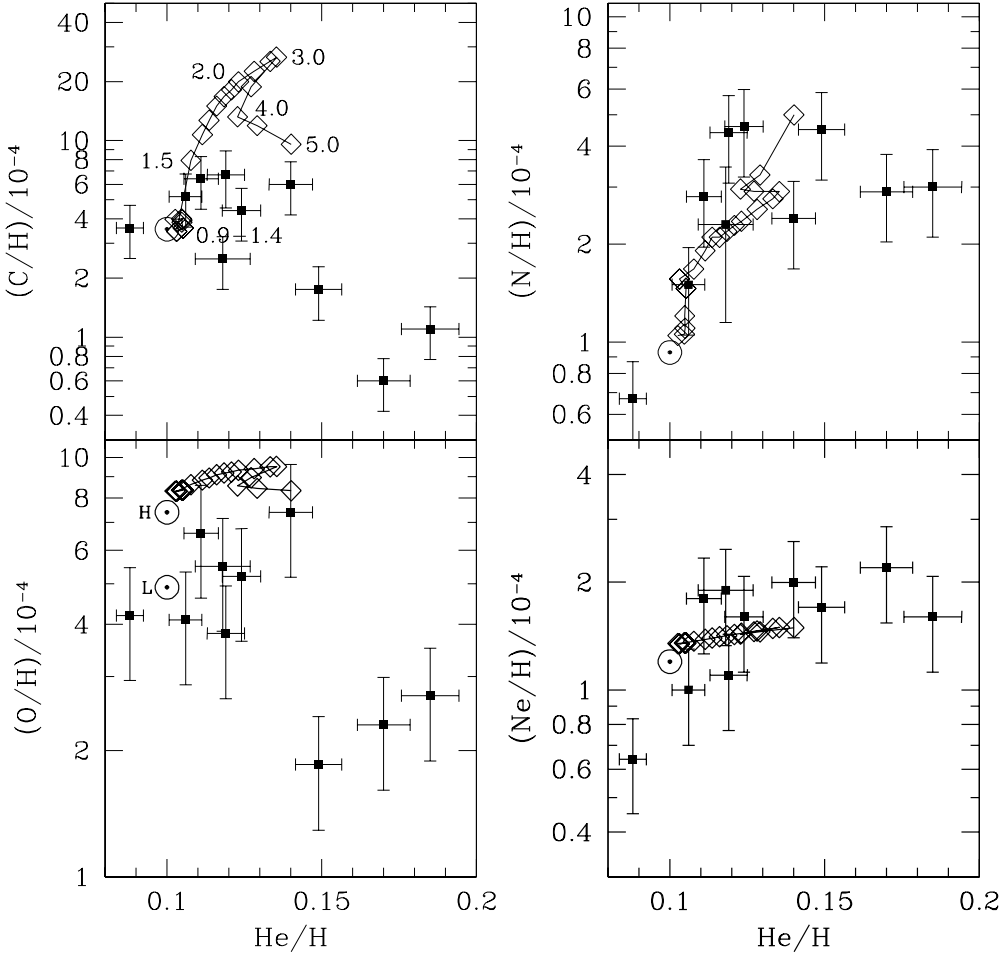}
\end{minipage}
\hfill
\begin{minipage}{0.21\textwidth}
\caption{Comparison between measured PN abundances   
(squares with errors bars) and model predictions (diamonds).
Two estimates for the oxygen abundance for the Sun 
are also indicated, O$_{\odot}$ High and O$_{\odot}$ Low
(see Table~{\protect{\ref{tab_mod}}} and text).
In each panel the sequence of diamonds denotes the expected 
PN abundance as a function  of the stellar progenitor's mass,
increasing from 0.9 to 5.0 $M_{\odot}$ 
(a few values are labeled along C/H curve), 
and for initial solar metallicity $Z=0.019$.
The initial O abundance was set equal to O$_{\odot}$ High.
The model assumptions adopted for the TP-AGB evolution,
corresponding to case A) of Table~{\protect{\ref{tab_mod}}},
include  solar-scaled molecular opacities ($\kappa_{\rm fix}$),
and third dredge-up efficiency $\lambda=0.5$ 
}
\label{fig1_z02kfix}
\end{minipage}
\end{figure*}

\section{Synthetic TP-AGB calculations }
\label{sec_tpagb}
In order to interpret the abundance data reported in Sect.~\ref{sec_obsab}, 
synthetic evolutionary calculations of the TP-AGB phase 
have been carried out with the aid of the model developed by Marigo et al. 
(1996, 1998, 1999), Marigo (1998, 2001, 2002), to whom we refer for all 
details.

We briefly recall that the initial conditions at the first thermal
pulse are extracted from full stellar models with  
convective overshooting by Girardi et al. (2000). 
The considered mass interval ($0.7 - 5.0\, M_{\odot}$) covers the whole 
class of low- and intermediate-mass stars that develop an electron-degenerate 
C-O core at the end of the core He-burning phase.
In these models the initial chemical composition is taken 
from  Anders \& Grevesse (1989) for solar metallicity ($Z=0.019$), 
while for other metallicities solar-scaled abundances are assumed.

The Girardi et al. (2000) models predict the changes (if any) in the surface 
chemical composition occurring prior to the TP-AGB phase by 
the first dredge-up during the first 
settling on the Red Giant Branch, and by the second dredge-up 
in stars of intermediate mass (say $M > 3.5-4\, M_{\odot}$) during the Early
AGB phase. 

The subsequent TP-AGB evolution is calculated from 
the first thermal pulse up to the complete ejection of the envelope 
by stellar winds. Mass loss is included according to the Vassiliadis \& Wood's 
(1993) formalism. The TP-AGB model predicts 
the  changes in the surface chemical composition 
caused by the third dredge-up and HBB. 
The third dredge-up is parametrized as a function of the efficiency, 
$\lambda$, and the minimum temperature at the base of the convective envelope, 
$T_{\rm b}^{\rm dred}$, required for dredge-up to occur
(Marigo et al. 1999, see also Sect.~\ref{ssec_modpre}).
The process of HBB -- expected to take place in the most massive and 
luminous AGB stars ($\ga 4.5 M_{\odot}$ depending on metallicity) --  
is followed  in detail with the aid of a complete envelope model
including the main nuclear reactions of the CNO cycle
(Marigo et al. 1998, Marigo 1998).
  
Recently Marigo (2002) has introduced a major novelty in the TP-AGB model,
that is the replacement of fixed solar-scaled molecular opacities
 -- commonly  adopted in most AGB evolution codes 
($k_{\rm fix}$) --
with variable molecular  opacities ($k_{\rm var}$) which are consistently
coupled with the actual elemental abundances of the outer stellar layers.
The impact of this new prescription on the evolution of AGB stars is 
significant and consequently, as will be shown below, it 
importantly affects the predictions of the PN elemental abundances. 

For the sake of clarity, in the following we will 
provide an outline of the main input assumptions adopted
in our TP-AGB calculations (Sect.~\ref{ssec_modpre} and Table~\ref{tab_mod}).  

\subsection{
Nucleosynthesis and mixing assumptions of the TP-AGB model}
\label{ssec_modpre}
The treatment of the third dredge-up in the TP-AGB
model is characterized by: 
\begin{itemize}
\item A temperature criterion to establish whether  
a dredge-up episode does or does not occur. It is based 
on the parameter $T_{\rm b}^{\rm dred}$, that corresponds to 
the  minimum temperature -- at the base of the convective envelope 
at the stage of post-flash luminosity maximum -- required for dredge-up
to take place.
In practice a procedure, based on envelope integrations, 
allows one to determine the onset of the third dredge-up, 
that is the minimum core mass $M_{\rm c}^{\rm min}$, and luminosity
at the first mixing episode.
More details can be found in Marigo et al. (1999; 
see also Wood 1981, Boothroyd \& Sackmann 1988, Karakas et al. 2002). 
\item The efficiency $\lambda = \Delta M_{\rm dred}/\Delta M_{\rm c}$.
It is defined as the fraction of the core mass increment, 
$\Delta M_{\rm c}$, over a quiescent inter-pulse period, 
that is dredged-up to the surface at the next thermal pulse
(corresponding to a mass $\Delta M_{\rm dred}$).
\item The composition of the convective inter-shell 
(in terms of the elemental abundances $X_{{\rm csh},i}$)
developed at thermal pulses, i.e. of the dredged-up material. 
We essentially specify the abundances of helium $X_{\rm csh}(^{4}$He),
carbon $X_{\rm csh}(^{12}$C), and oxygen $X_{\rm csh}(^{12}$O)
(see Table~\ref{tab_mod}).
We recall that detailed calculations of thermal pulses indicate
a typical inter-shell composition [$X_{\rm csh}(^{4}$He$) \sim 0.76$;
$X_{\rm csh}(^{12}$C$) \sim 0.22$; $X_{\rm csh}(^{16}$O$) \sim 0.02$]
(see e.g. Boothroyd \& Sackmann 1988; Forestini \& Charbonnel 1997).
However, these results may drastically change with the inclusion
of extended overshooting from convective boundaries 
(i.e. $X_{\rm csh}(^{12}$C$) \sim 0.50$; 
$X_{\rm csh}(^{16}$O$) \sim 0.25$ according to Herwig et al. 1997),
or in the most massive TP-AGB models with deep dredge-up penetration
(i.e. Vassiliadis \& Wood 1993; Frost et al. 1998).
This latter possibility, relevant to our analysis, 
is discussed in Sect.~\ref{sssec_cn}.  
 
In addition to $^{4}$He, $^{12}$C, and $^{16}$O, 
we account for the possible production of $^{22}$Ne, 
via the chain of reactions $^{14}$N($\alpha$, $\gamma$)\, 
$^{18}$F($\beta^+$, $\nu$)\,
$^{18}$O($\alpha$, $\gamma$)\, $^{22}$Ne (Iben \& Renzini 1983).
Then a certain amount of $^{22}$Ne may be burned via the neutron source 
reaction $^{22}$Ne($\alpha$, n)$^{25}$Mg. The efficiency of this nuclear step
strongly depends on the maximum temperature achieved 
at the bottom of the inter-shell developed at thermal pulses,
being marginal  for $T < 3 \times 10^{8}$ K 
(about 1 $\%$ of $^{22}$Ne is converted 
into $^{25}$Mg), while becoming more
and more important at higher temperatures (Busso et al. 1999).

In our study, we parameterize the Ne and Mg production assuming  
that the abundances
in the convective intershell are given by (see also Marigo et al. 1996):
\begin{equation}
X_{\rm csh}(^{22}{\rm Ne}) =\,X_{\rm e}(^{22}{\rm Ne}) + 
F \times \, \displaystyle{\frac{22}{14}} \times X_{\rm Hsh}(^{14}{\rm N}) 
\label{eq_ne22}
\end{equation}
\begin{displaymath}
X_{\rm csh}(^{25}{\rm Mg}) =\, X_{\rm e}(^{25}{\rm Mg})  + 
(1-F) \times \displaystyle{\frac{25}{14}} \times \, X_{\rm Hsh}(^{14}{\rm N}) 
\end{displaymath}
where $X_{\rm e}$ refers to the envelope abundance before the dredge-up,
and $F$ represents the degree of efficiency of the $^{22}$Ne production,
that is clearly complementary to that of $^{25}$Mg ($1-F$).
In most of our calculations we assume $F = 0.99$, that means allowing 
the chain of $\alpha$-capture reactions to proceed from 
$^{14}$N to $^{22}$Ne, with essentially no further production of $^{25}$Mg.
This point is discussed in Sect.~\ref{sssec_ne}.

In the above equations  $X_{\rm Hsh}(^{14}{\rm N})$ denotes 
the nitrogen abundance left by the H-burning 
shell in the underlying inter-shell region, just before the occurrence
of the pulse. It is estimated  with
\begin{displaymath}
\begin{array}{ll}
X_{\rm Hsh}(^{14}{\rm N}) = 14 \times 
\displaystyle{\left[
\frac{X_{\rm e}(^{12}{\rm C})}{12}+ 
\frac{X_{\rm e}(^{13}{\rm C})}{13}+  
\frac{X_{\rm e}(^{14}{\rm N})}{14} + \right.} \\
\displaystyle{\left.
\frac{X_{\rm e}(^{15}{\rm N})}{15}+
\frac{X_{\rm e}(^{16}{\rm O})}{16}+ 
\frac{X_{\rm e}(^{17}{\rm O})}{17}+ 
\frac{X_{\rm e}(^{18}{\rm O})}{18} \right]} \, ,\\
\end{array}
\end{displaymath}
i.e. we assume that all CNO nuclei are converted in  $^{14}{\rm N}$, which is
a good approximation when the CNO-cycle operates under equilibrium
conditions.
In this way we account for the possible primary component of $^{14}{\rm N}$
in the inter-shell,  which is produced every time  some
freshly dredged-up $^{12}{\rm C}$ in the envelope is engulfed 
by the H-burning shell during the quiescent inter-pulse evolution.  

As a consequence, the resulting  $X_{\rm csh}(^{22}{\rm Ne})$, 
synthesized via the chain of Eq.~(\ref{eq_ne22}),  
contains a primary component, which can also be injected into 
the surface chemical composition through the dredge-up.
We note that $^{22}{\rm Ne}$ is expected to be purely  secondary
in low-mass stars that do not experience the third dredge-up.

 
\end{itemize}

The reader can find more details about 
model prescriptions in Table~\ref{tab_mod}. 

\begin{figure*}
\begin{minipage}{0.77\textwidth}
\includegraphics[width=\textwidth]{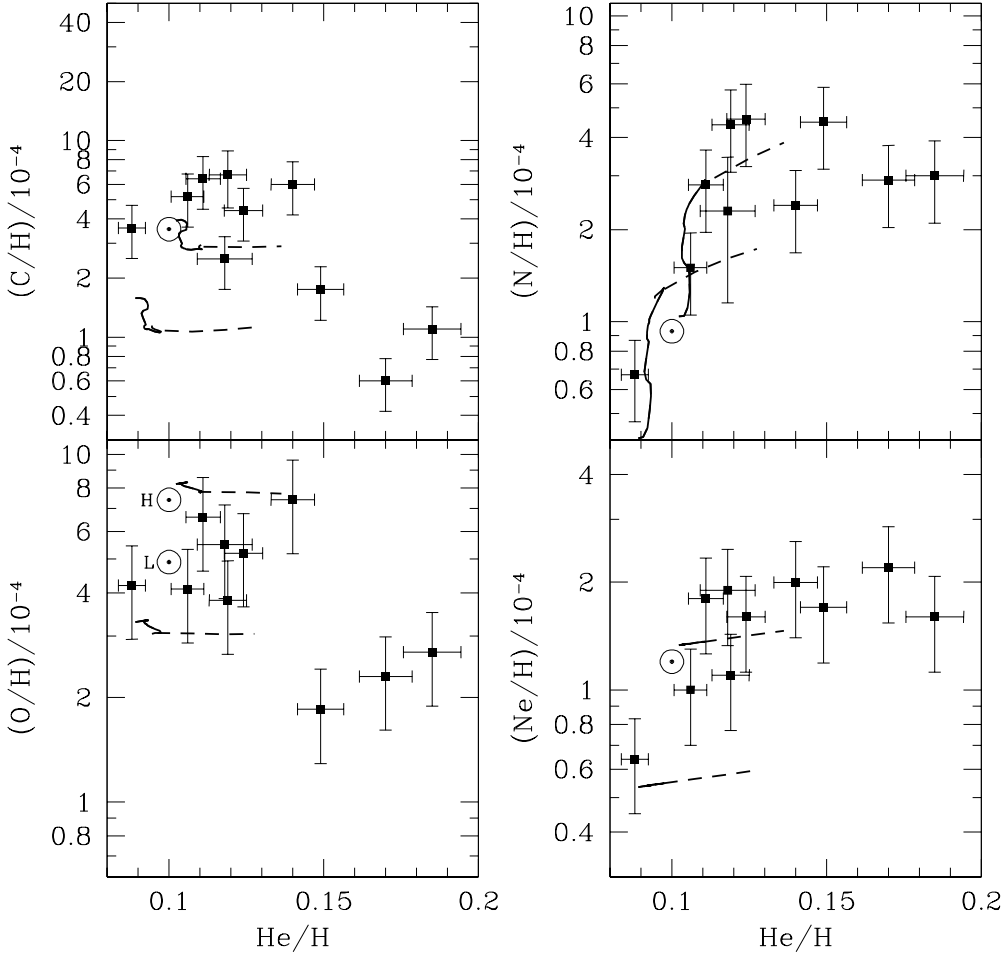}
\end{minipage}
\hfill
\begin{minipage}{0.21\textwidth}
\caption{Observed PN abundances 
(squares with error bars)
compared to expected surface abundances (taken from Girardi et al. 2000) 
just before the onset of the TP-AGB phase, as a function of the 
initial stellar mass, and for two values of the original metallicity, 
i.e. $Z=0.019$ (upper curves) and $Z=0.008$ (lower curves).
The initial mass ranges from 0.6 to 5.0 $M_{\odot}$ 
at increasing  He/H along the curves.
Predicted abundances are those expected after  
the first dredge-up (solid line) for stars with $0.6 \le M/M_{\odot} \le 4.0$;
and after the second dredge-up (dashed line) for stars with 
$4.0 < M/M_{\odot} \le  5.0$ 
}
\label{pn1_z02_d12}
\end{minipage}
\end{figure*}

\section{Comparison between models and observations}
\label{sec_obsmod}

We will now perform an analysis of the observed PN elemental abundances 
with the aid of the TP-AGB models just described.
The basic idea is to constrain the  model parameters 
so as to reproduce the observed data, and hence derive 
indications on the evolution
and nucleosynthesis of  AGB stellar progenitors.
The chemical elements under consideration are He, C, N, O and Ne.

\subsection{The starting point: TP-AGB models with solar-scaled 
molecular opacities}
\label{ssec_kfix}
First ``old'' predictions of
PN abundances (Marigo 2001) are considered; 
see  Fig.~\ref{fig1_z02kfix}.
In those models with an initially solar composition 
the dredge-up parameters 
($\lambda$ and $T_{\rm b}^{\rm dred}$) were calibrated 
to reproduce the observed carbon star luminosity functions in 
both Magellanic Clouds (Marigo et al. 1999). 
Moreover, envelope integrations were carried out using 
fixed solar-scaled molecular opacities 
($\kappa_{\rm fix}$; see Sect.~\ref{sec_tpagb}).
   
By inspecting Fig.~\ref{fig1_z02kfix} we note that, 
though a general agreement is found 
between measured and predicted abundances with respect to 
nitrogen and neon abundances, 
three main discrepancy points occur:
\begin{enumerate}
\item A sizeable overproduction of carbon by models, up to a factor of 3-5;
\item The lack of extremely He-rich models, with 0.15 $<$ He/H $\le
  0.20$;
\item A general overabundance of oxygen in all models, amounting up 
to a factor of 3. 
\end{enumerate}

While the first two aspects are probably related to the nucleosynthetic
assumptions in the TP-AGB models, the third one could also
reflect our choice of oxygen abundance  
in the solar mixture (Anders \& Grevesse 1989), which has 
recently been subject of major revision (Allende Prieto et al. 2001).
The aim of the following analysis is to single out the main causes of 
the disagreement and  possibly to remove it by proper changes in the
model assumptions.

\subsection{Sub-grouping of the ISO sample and comparison with other PNe data}
%
%
  \begin{figure*}[]
  \begin{center}
    \begin{picture}(500,380)(0,0)
      \psfig{figure=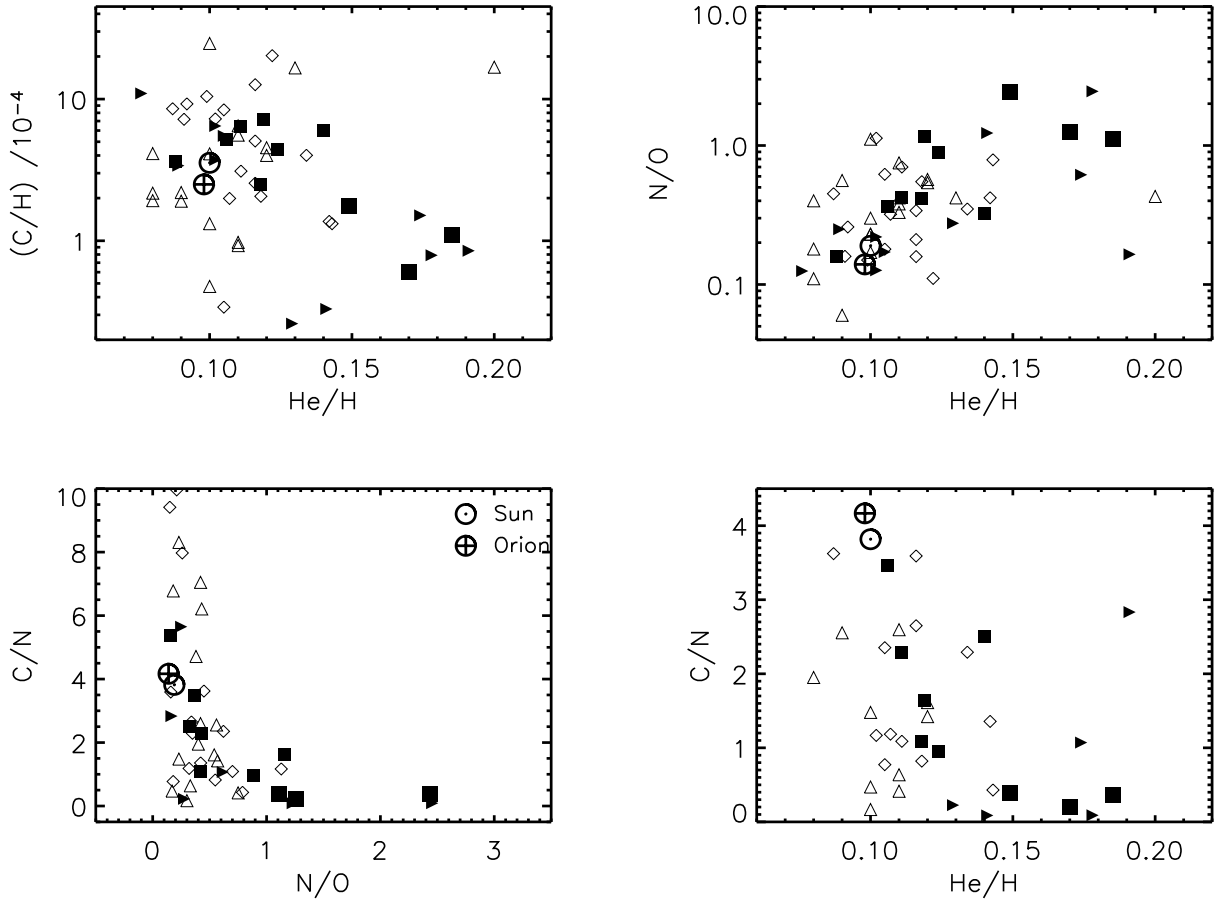,width=500pt,angle=90}
    \end{picture}
  \end{center}
\caption{Comparison of PN abundances from our ISO sample (filled squares) 
and other samples in the literature, namely: 
Henry et al. (2000; open triangles) and Kingsburgh \& Barlow (1994; diamonds) 
for PNe in the Milky Way, and Dopita et al. (1997; filled triangles) 
for PNe in the LMC.
The ISO sample has been plotted using smaller and larger squares for PNe
with helium abundances lower and higher than 0.145 respectively} 
\label{fig_comp}
\end{figure*}

Before starting the interpretative analysis, it 
is worth noticing that the PNe data (see Fig.~\ref{fig1_z02kfix}) 
seem to segregate in two different 
groups in the observational abundance plots -- one at lower (He/H $\leq$ 0.14)
and the other at higher helium content (He/H $>$ 0.14) --, 
and their chemical patterns
for carbon and nitrogen already suggest likely different 
mass ranges for the progenitors 
(i.e. low- and intermediate-mass stars  respectively).

In particular, there are three PNe, \object{He\,2-111}, \object{NGC\,6302} and
\object{NGC\,6537} that clearly show the highest
helium abundance (He/H $>$0.14) together with the lowest carbon and
oxygen abundances.
To this respect we recall that it was recognized a long time
ago (Becker \& Iben 1980) that the observed N/O-He/H correlation and
the C/O-N/O anti-correlation -- characterising the He-rich PNe --  
present a problem for models of nucleosynthesis on the AGB.
This will be discuss in Sect.~\ref{ssec_highe}.

Comparison with literature data on PN elemental abundances provides
further support for two separate groups of PNe according to their helium
content, ``normal'' or high.
Figure\,\ref{fig_comp} compares our observed abundances  with data from
Henry et al.  (2000) and Kingsburgh \& Barlow (1994) for PNe in the
Milky Way, and from Dopita et al. (1997) for PNe in the LMC.  Only PNe with
measured C abundances and not already present in the ISO sample 
have been included.  The reader should bear in
mind that excluding helium, the errors on the remaining abundances of these
two samples are large and uncertain. 

The ``normal'' helium group in the ISO sample 
(with He/H $<$ 0.145; smaller filled squares) 
is within the range of abundances
given by Henry et al.  (2000) and Kingsburgh \& Barlow (1994).  However,
both studies lack PNe with high He abundances.  
Actually, PN \object{PB\,6} (from
Henry et al. (2000) sample) has a helium abundance of 0.2 but that value
is highly uncertain, and 0.146 is the highest helium abundance 
among those PNe that we have excluded from the 
Kingsburgh \& Barlow (1994) sample
(those with no measured carbon).

Conversely,  
the three ISO PNe with high helium abundances 
(with He/H $>$ 0.145; larger filled squares)
agree well with the LMC PNe from Dopita et al. (1997), 
not only in the He abundance but also in the N/O and C/N ratios. 
Besides supporting the
division of the ISO sample into two subclasses, this also suggests that
an initial sub-solar metallicity may be a key factor in the subsequent
nucleosynthesis.

For all these reasons, it seems advantageous
to separate those PNe which exhibit high helium 
content from the others having ``normal'' helium   
abundances, and to discuss the two groups separately.

%
\begin{figure*}
\includegraphics[width=\textwidth]{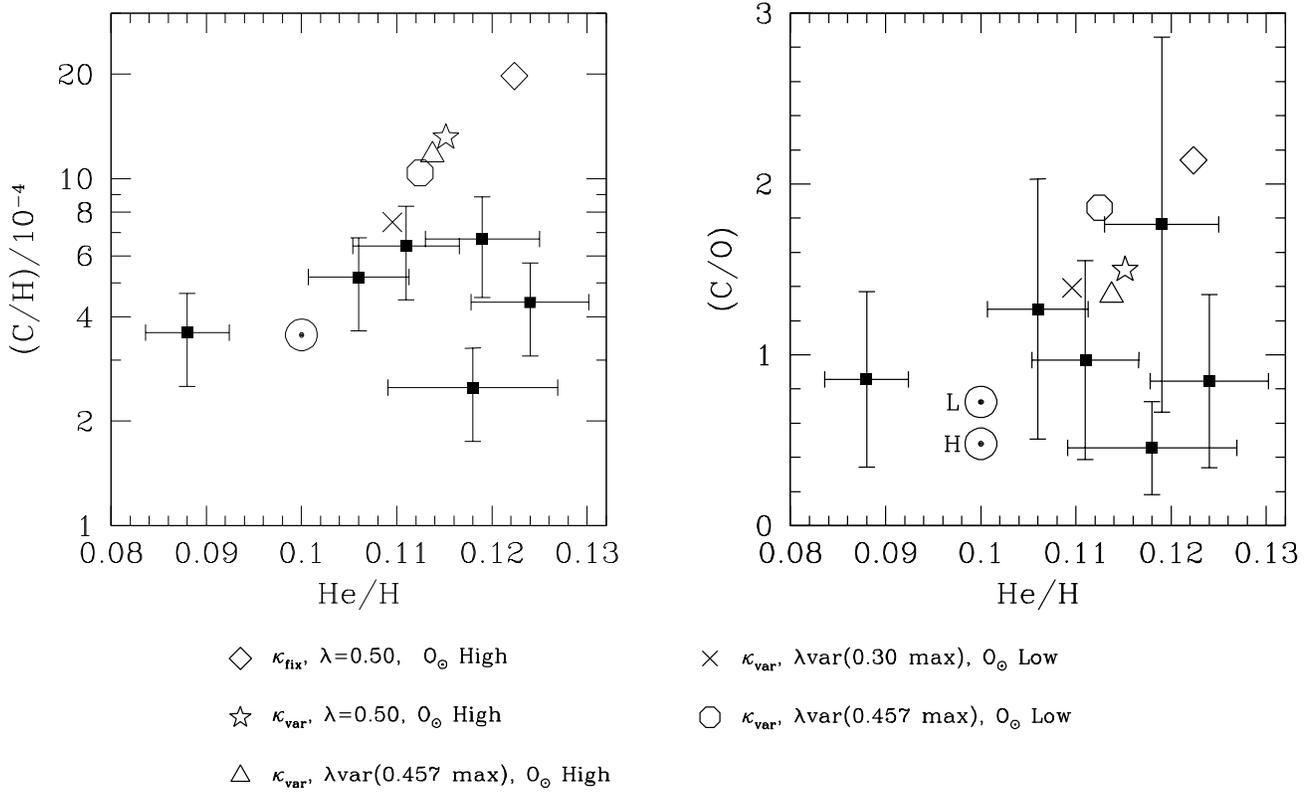}
\caption{Carbon abundances and  C/O ratios in PNe with 
low helium abundances (He/H$\le 0.13$). 
Open symbols denote predictions of synthetic TP-AGB calculations 
for various choices of model parameters -- specified in the legenda -- 
corresponding to a stellar progenitor with  
$(M_{\rm i} = 2.0 M_{\odot}, \, Z_{\rm i} = 0.019)$. 
See text and cases B-C-D-E of Table~{\protect{\ref{tab_mod}}} for more details 
} 
\label{fig_che2}
\end{figure*}

\begin{figure*}
\begin{minipage}{0.77\textwidth}
\includegraphics[width=\textwidth]{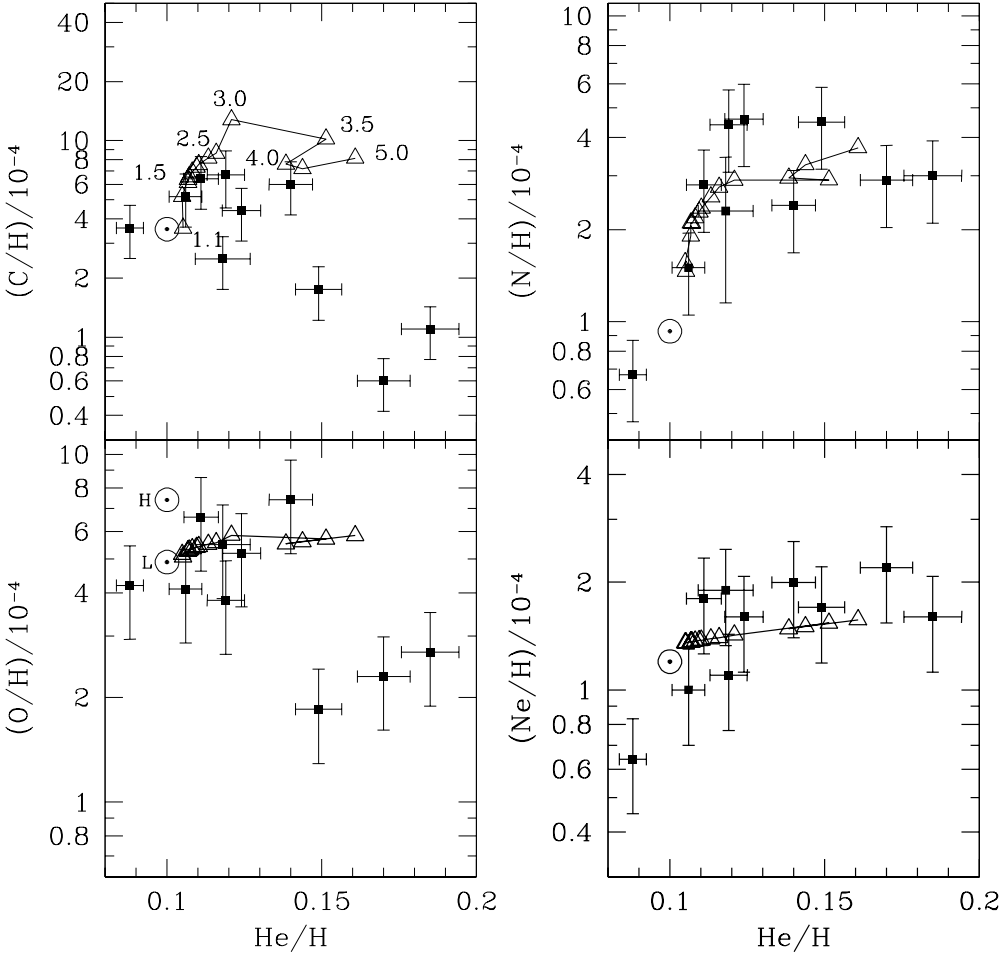}
\end{minipage}
\hfill
\begin{minipage}{0.21\textwidth}
\caption{Summary of the results that best fit the 
observed PN abundances for He/H$\le 0.14$.
Specifically, triangles represent predictions
derived from  TP-AGB models with $Z=0.019$ and 
with initial stellar masses from 1.1 to 5.0 $M_{\odot}$, and adopting the 
following prescriptions summarised in Table~\ref{tab_mod}:
case E) for $1.1 \le M/M_{\odot} \le 2.5$;
case D) for $M = 3.0\, M_{\odot}$;
case F) for $3.5 \le M/M_{\odot} \le 5.0$.
In practice, we assume the efficiency of the third dredge-up 
changes during the evolution (according to Karakas et al. 2002), 
and increases with the stellar mass. Note that also the composition 
of the convective inter-shell varies as a function of the stellar mass,
i.e. less primary carbon is supposed to be synthesised during thermal 
pulses by models with the largest masses.
See text for more details}
\label{fig1_z02lowmkvar}
\end{minipage}
\end{figure*}
 
\begin{figure*}
\includegraphics[width=\textwidth]{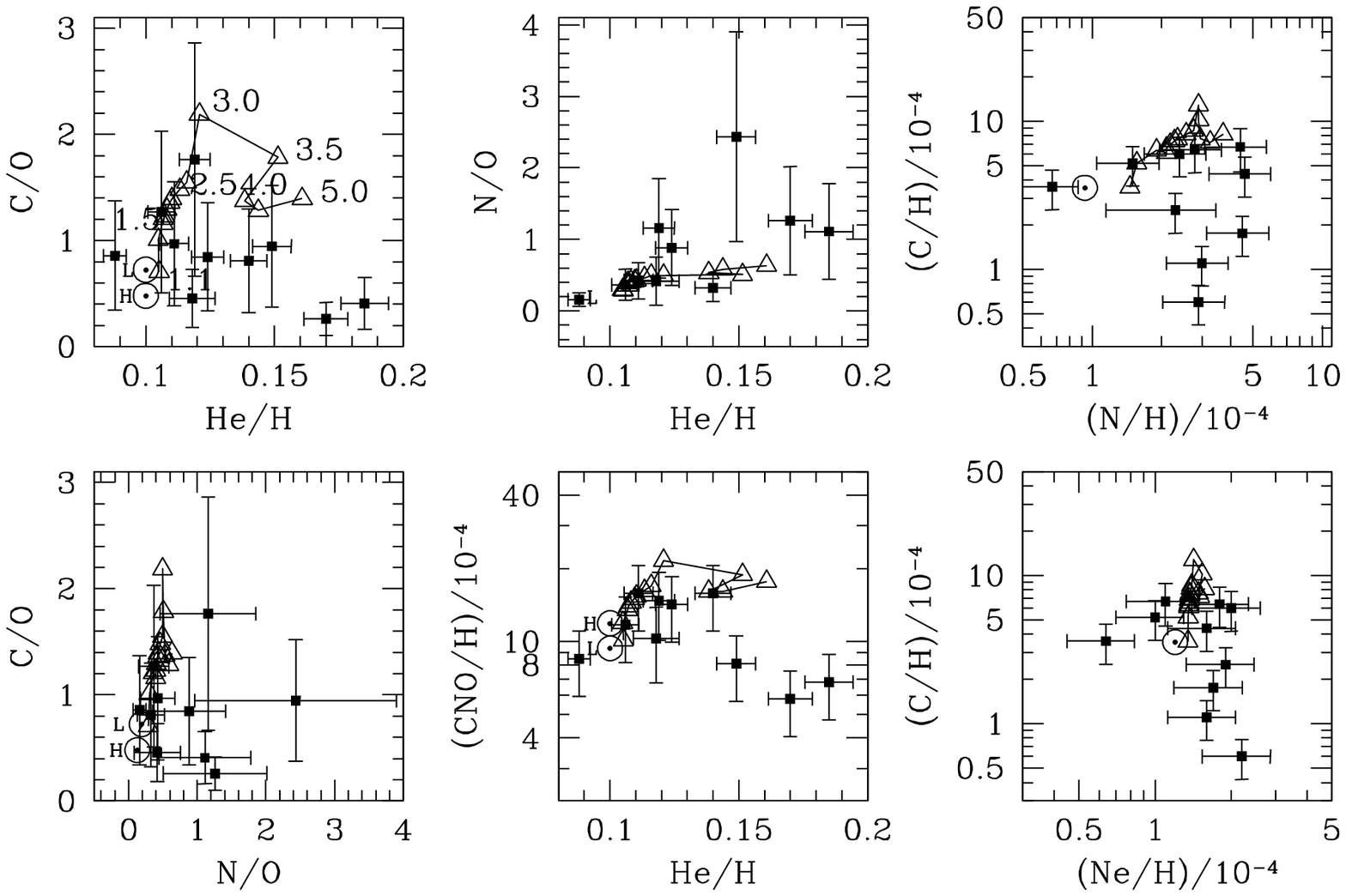}
\caption{The same as in Fig.~\ref{fig1_z02lowmkvar},
but expressing the abundance data with different combinations 
of elemental ratios}
\label{fig2_z02lowmkvar}
\end{figure*}


\subsection{PNe with ``normal'' He abundances}
\label{ssec_normhe}
Let us examine the PN data with  He/H$ \la 0.14$
(Fig.~\ref{fig1_z02kfix}).
For the sake of clarity we summarise
the main features pointed out in Sect.~\ref{sec_obsab}. 
\begin{itemize}
\item Most of the data present a clear enhancement in He, C, N compared to 
the abundances of these elements for the Sun. In particular, 
some of them should 
descend from carbon stars  given their C/O ratio larger than one. 
\item Oxygen abundances are underabundant when compared to the Solar
determination by Anders \& Grevesse (1989). On the other hand they are 
consistent with a constant, or slightly increasing trend with He/H, 
if compared 
to the recent oxygen estimation for the Sun by Allende Prieto et al. (2001).
\item Neon seems to exhibit a constant, or perhaps moderately increasing trend
 with He/H. 
\end{itemize}

We first consider the elemental changes expected after the first and second 
dredge-up episodes. The predicted envelope abundances are displayed 
in Figure~\ref{pn1_z02_d12} for two values  of the initial metallicity, 
i.e. $Z=0.019$ and $Z=0.008$.
Compared to their original values oxygen and neon are essentially unaffected, 
carbon is somewhat reduced, while nitrogen and helium may be significantly 
increased particularly after the  second dredge-up in stars 
of intermediate mass ($M > 4\, M_{\odot}$).

The comparison with the  PN data indicates   
the necessity to invoke further chemical changes 
in addition to those caused by the first and second dredge-up, especially
if one considers the observed enhancement in carbon.
The most natural explanation resides in the third dredge-up process 
occurring during the TP-AGB phase. To this respect,  as already mentioned
in Sect.~\ref{ssec_kfix}, the ``old'' models with fixed solar-scaled molecular
opacities predict too large carbon enrichment (Fig.~\ref{fig1_z02kfix}).
This point and its possible solution are discussed below.

\subsubsection{Reproducing the extent of carbon enrichment} 
To tackle the problem of carbon we consider a stellar model with 
$2\, M_{\odot}$ initial mass,  a typical 
mass of carbon stars in the Galactic disk (Groenewegen et al. 1995).  
Figure~\ref{fig_che2} shows a few models of the predicted
carbon abundance and corresponding PN values.

As starting model we consider the one with the $\kappa_{\rm fix}$ 
assumption. It is clearly located well above the observed points 
(diamond in left panel of Fig.~\ref{fig_che2}).

A first significant improvement is 
obtained acting on the opacity,
that is adopting variable molecular opacities during the TP-AGB 
calculations. We replace  
$\kappa_{\rm fix}$ with $\kappa_{\rm var}$, while keeping
the other model parameters  fixed.
In this model (starred symbol) the predicted C/H is lowered because 
of a shorter  C-star phase, hence a decrease 
in the  number of dredge-up episodes (see Marigo 2002, 2003).

Additional cases are explored.
The usual assumption of constant dredge-up efficiency $\lambda$
is replaced with the recent recipe by Karakas et al. (2002),
who provide analytic relations -- fitting the results of full AGB
calculations -- that express  the evolution of $\lambda$ as a function
of metallicity $Z$, stellar mass $M$, and progressive pulse number. 
For a given $M$ and $Z$, 
$\lambda$ is  found to increase from initially zero up to nearly 
constant maximum value, $\lambda_{\rm max}$. This latter also varies
depending on  mass loss.
Adopting  $\lambda_{\rm max}= 0.457$ as suggested by Karakas et al. (2002)
for the ($M=2\, M_{\odot}, Z=0.02$) combination, we end up with a somewhat
lower C/H (triangle), compared to the case with 
($\kappa_{\rm var}$, $\lambda=0.5$). 
This reflects the smaller  amount of carbon globally dredged-up with 
the $\lambda_{\rm var}$ assumption.  

A further test is made  with respect to the assumed initial oxygen abundance.
We calculate the TP-AGB evolution 
of the $2\, M_{\odot}$ model replacing the surface oxygen abundance 
at the first thermal pulse -- as predicted by Girardi et al. (2000)
evolutionary calculations, based on Anders \& Grevesse (1989) solar
mixture (the O$_{\odot}$ High in Table \ref{tab_mod}) --  
by the recently revised determination by Allende Prieto et al. 
(2001; the O$_{\odot}$ Low in Table \ref{tab_mod}).

With the new lower oxygen abundance the 
solar C/O ratio increases from 0.48 to 0.78.
This increment in the initial C/O ratio has important effects on the 
later TP-AGB evolution and PN abundances: fewer dredge-up episodes are 
necessary to produce a carbon star and hence, on average, 
a lower C abundance is expected in the PN ejecta. 
This can be seen in Fig.~\ref{fig_che2}, by comparing the models
labeled by  triangle and circle symbols.
We also notice that the trend in the predicted values of  C/H 
(left panel) or C/O (right panel) is reversed.
The O$_{\odot}$ Low prescription leads to a reduced absolute carbon 
enrichment compared to the O$_{\odot}$ High case, while the final 
C/O ratio is larger in the former case. 
 
At this point the results already appear better 
compared to the starting model, but in order to  carry the expected 
C/H point within the observational error
bars, we change another fundamental model parameter, i.e. the
dredge-up efficiency $\lambda$.
A good fit of the PNe data is obtained by lowering $\lambda_{\rm max}$ 
from 0.457 to 0.3, that simply means diminishing the amount of 
dredged-up carbon. 

In summary, from these calculations we derive the
following indications. Both the C/H and the C/O
ratios of the carbon-rich Galactic PNe --
evolved from stars with typical masses of $\sim 2.0\, M_{\odot}$ -- 
can be reproduced by assuming
i) variable molecular opacities  $\kappa_{\rm var}$, ii) an initial
oxygen abundance as recently revised by Allende Prieto et al. (2001), and 
iii) a dredge-up efficiency $\lambda \approx 0.3-0.4$.
These indications -- 
in particular points i)  and iii) 
-- agree with the results recently obtained by  Marigo (2002, 2003) in her
analysis of the properties of Galactic carbon stars in the disk.
Variable molecular opacities and $\lambda \la 0.5$ are required to 
reproduce a number of observables of carbon stars, like their C/O ratios,
effective temperatures, mass-loss rates, and near-infrared colours. 

Then, from the representative case of the  $2.0\,  M_{\odot}$ model, 
we consider a wider mass range, i.e. $1.1-5.0\, M_{\odot}$, 
with initial metallicity $Z=0.019$.
Results of the  C/H and C/O ratios as a function
of He/H, expected in the corresponding PNe, 
are displayed in Figs.~\ref{fig1_z02lowmkvar} and \ref{fig2_z02lowmkvar} and are also summarized in the top-left panels of Figs.~\ref{fig1_z02kvar} 
and \ref{fig2_z02kvar}
(triangles connected by solid line).
They show a satisfactory agreement with 
the observed data for He/H$<0.14$, suggesting a
mass interval for the stellar progenitors from  about 1.0 to 4.0 $M_{\odot}$

Synthetic TP-AGB calculations are carried out by adopting the 
$\kappa_{\rm var}$ and O$_{\odot}$ Low prescriptions, and  
assuming $\lambda \approx 0.3-0.5$ for 
lower stellar masses ($M \la 3.0\, M_{\odot}$), while increasing it
up to $\lambda \approx 0.98$ for the largest masses. 
This means that the third dredge-up should become deeper 
in more massive stars, as 
indicated by full TP-AGB calculations (e.g. Vassiliadis \& Wood 1993;
Karakas et al. 2002).
It is worth noticing that models with $M \le 1.5\, M_{\odot}$  are predicted
not to experience the third dredge-up, i.e. they do not fulfil the
adopted temperature criterion  (based on the parameter 
$T_{\rm b}^{\rm dred}$; refer to Sect.~\ref{ssec_modpre}). 

 Moreover, we assume that in the most massive stars
(with $M = 4.0-5.0\, M_{\odot}$) the composition of the inter-shell 
may be different from the standard one, i.e. less primary carbon 
is supposed to be synthesised during thermal pulses. This point will be
discussed in more detail in Sect.~\ref{sssec_cn}.

\subsubsection{Other elemental abundances}
As for the He, N, O, Ne elemental abundances, we can outline the following
(see sequences of triangles in Figs.~\ref{fig1_z02kvar}, \ref{fig2_z02kvar}):
\begin{itemize}
\item The helium abundance up to He/H$\sim 0.14$ is well 
reproduced by accounting for the 
surface enrichment due to the first, and possibly second and third dredge-up.
Apparently, for this group of PNe,  
there is no need to invoke a significant production of
helium by HBB. We also notice that the minimum value predicted by 
stellar models with initial solar composition is He/H$\sim 0.096$, so that 
any observed value lower than that (e.g. \object{NGC\,7662}) 
may correspond to a stellar progenitor with lower 
initial metallicity and helium content. 
\item The nitrogen data are consistent, on average, with 
the expected enrichment
produced by the first and second dredge-up events.
\item As already discussed, the PN oxygen estimates are compatible with 
the recent revision for the abundance in the Sun
by Allende Prieto et al. (2001; $({\rm O/H}_{\odot}) = 4.9 \times 10^{-4}$). 
The PN data indicate that 
during the evolution of the stellar progenitors, 
their surface abundance of oxygen is essentially unchanged, 
or it might be somewhat enhanced.      
Anyhow, the revised lower determination for the Sun
removes the problem of explaining the systematic 
oxygen under-abundance compared to solar that 
all data would present if adopting  higher values for the Sun 
as indicated by past analyses (e.g. Anders \& Grevesse 1989; 
$({\rm O/H})_{\odot} = 7.4 \times 10^{-4}$).  
\item 
The neon data do not allow to put stringent constraints on the
nucleosynthesis of this element in TP-AGB stars with initial solar metallicity.
In fact, within the uncertainties the observed Ne/H abundances  
are compatible with  a 
preservation of the original value, but they can also point to 
a moderate increase.

Predictions shown in Fig.~\ref{fig1_z02kvar} are obtained under the 
assumption that the synthesis 
of $^{22}$Ne -- via $\alpha$-captures in the He-burning shell 
starting from $^{14}$N during thermal pulses -- 
takes place with almost maximum possible efficiency 
($F=0.99$, see Sect.~\ref{ssec_modpre}).
As we see, the final expected enrichment in PNe remains modest,
due to the relatively  small number of thermal pulses and 
moderate dredge-up efficiency ($\lambda \sim 0.3-0.5$) characterising 
TP-AGB models with initial masses $M \sim 1.5-3.0$, and metallicity 
$Z=0.019$.

The opposite situation, that is the complete conversion of $^{22}$Ne 
into $^{25}$Mg with $F \sim 0$, would not lead to any  
enrichment in  neon so that 
the sequence of predicted Ne/H  would be
flatter\footnote{In any case Ne/H should keep a slightly increasing
trend, not becoming exactly horizontal since,  
even if Ne$=const.$, the hydrogen content H decreases as a 
consequence of dredge-up events.}  
than the present one (connected triangles in Fig.~\ref{fig1_z02kvar}).
But this case could not be rejected either since corresponding predictions 
are still confined within the observed domain. 


\end{itemize}

\subsection{PNe with extremely high He abundance}
\label{ssec_highe}
Below we discuss the PNe exhibiting 
the largest enrichment in helium, i.e. with $0.14  \la$ He/H$ \la
0.20$ (see e.g. Fig.~\ref{fig1_z02kfix}). These objects
share other chemical properties, namely:
\begin{itemize}
\item Marked carbon deficiency compared to the solar value;
\item Sizeable enhancement of nitrogen, but not exceeding the 
upper values observed in the PNe  with lower He/H;
\item Significant depletion of oxygen, which makes these PNe
appear as a distinct group with respect to the
PNe with lower He/H abundances; 
\item Possible, but still not compelling, hint of  
over-abundance of neon compared to the solar value.
\end{itemize}
\begin{figure*}
\begin{minipage}{0.77\textwidth}
\includegraphics[width=\textwidth]{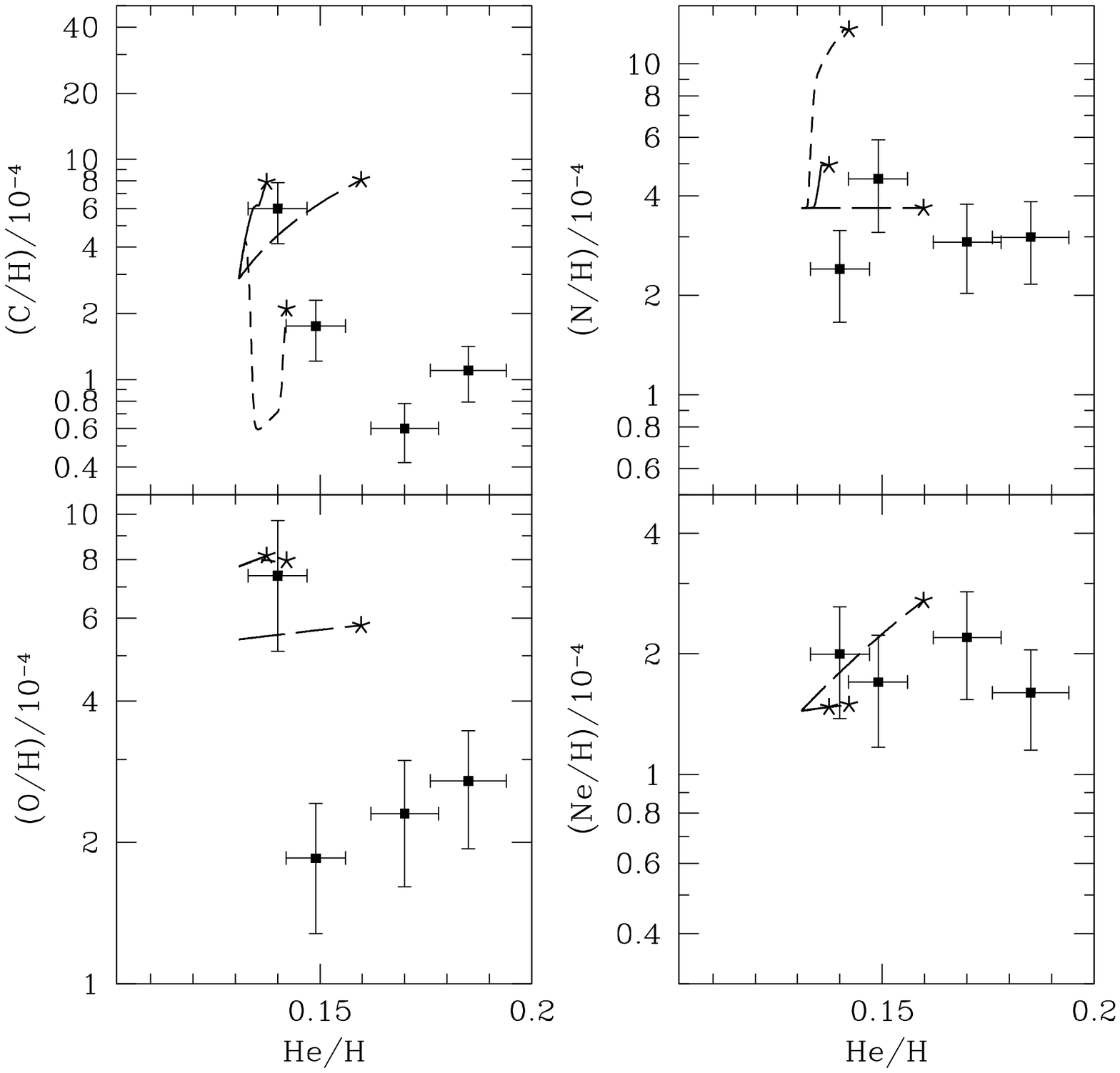}
\end{minipage}
\hfill
\begin{minipage}{0.21\textwidth}
\caption{Time evolution of surface elemental abundances during the
TP-AGB phase of a $5.0 M_{\odot}$ model with initial solar
chemical composition, experiencing both the third
dredge-up and HBB. 
Observed PN data should be compared with the starred symbol 
at the end of each curve (marking the end of the TP-AGB phase).
Most parameter prescriptions are specified 
Table~\ref{tab_mod}. In practice
we consider the following cases: 
i) efficient HBB with $\alpha=2.50$ 
(short-dashed line; refer to case A) of Table~\ref{tab_mod} for other model
parameters);
ii) weak HBB with $\alpha=1.68$ (solid line; refer to case A) 
of Table~\ref{tab_mod}); 
and iii) weak HBB and efficient dredge-up, starting from a 
lower oxygen abundance (long-dashed line; refer to case F) 
of Table~\ref{tab_mod})
}
\label{fig_hbbz02}
\end{minipage}
\end{figure*}
\begin{figure*}
\begin{minipage}{0.77\textwidth}
\includegraphics[width=\textwidth]{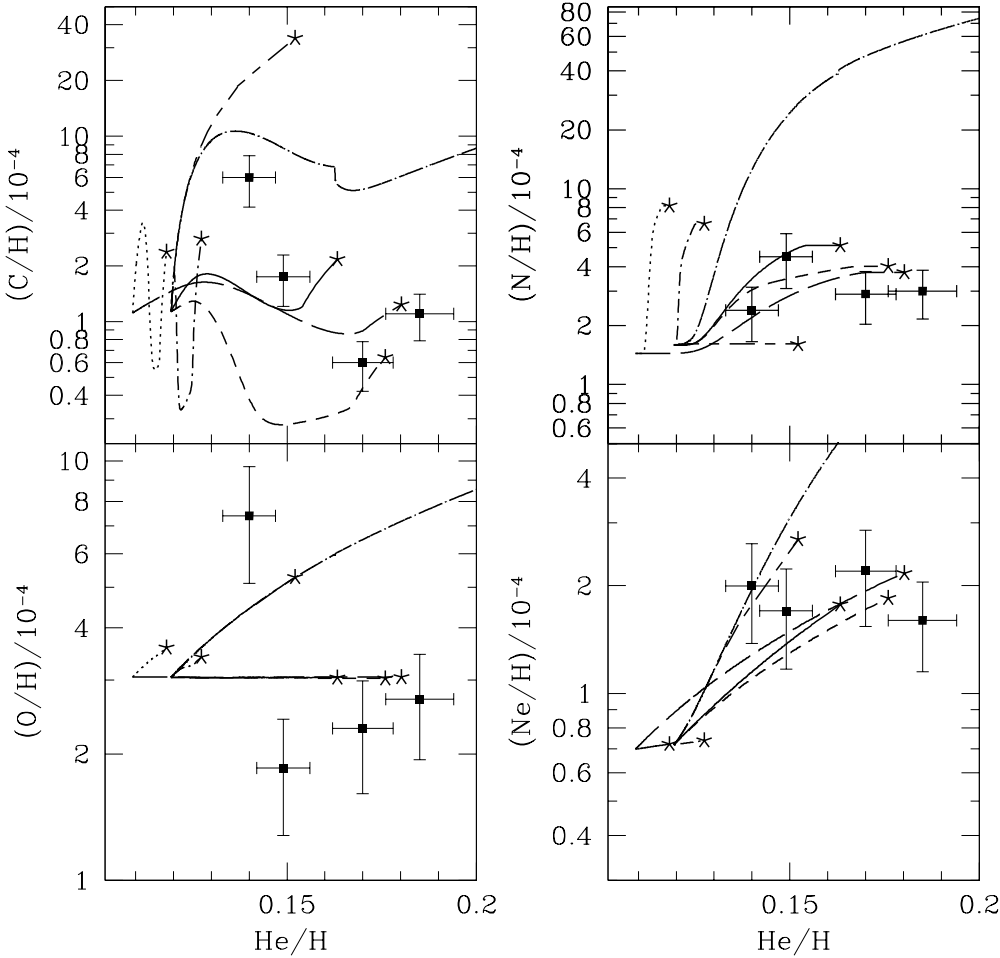}
\end{minipage}
\hfill
\begin{minipage}{0.21\textwidth}
\caption{The same as in Fig.~\protect{\ref{fig_hbbz02}}, but for
initial LMC chemical composition. 
Illustrated cases mainly differ in the adopted parameters describing  
the third dredge-up, i.e. (see also Table~\ref{tab_mod}):
i) and ii) moderate third dredge-up ($\lambda=0.5$) 
and standard chemical composition of the inter-shell (case G) 
for $4.5$ and $5.0\, M_{\odot}$  models 
(dotted and dot-short-dashed lines, respectively),
iii) deep third dredge-up ($\lambda=0.9$) and standard chemical composition 
of the inter-shell (case H) for  $5.0\, M_{\odot}$
model (dot-long-dashed line);
iv) the same as the previous case but for the $\kappa_{\rm var}$
prescription (case I; short-long-dashed line);
v), vi), and vii) deep third dredge-up ($\lambda=0.9$) and inter-shell
chemical composition with low-carbon abundance  for 
$5.0\, M_{\odot}$ model (solid line and case J;  short-dashed line and case K)
and $4.5\, M_{\odot}$ model (long-dashed line and case K)
}
\label{fig_hbbz008}
\end{minipage}
\end{figure*}
\begin{figure*}
\begin{minipage}{0.77\textwidth}
\includegraphics[width=\textwidth]{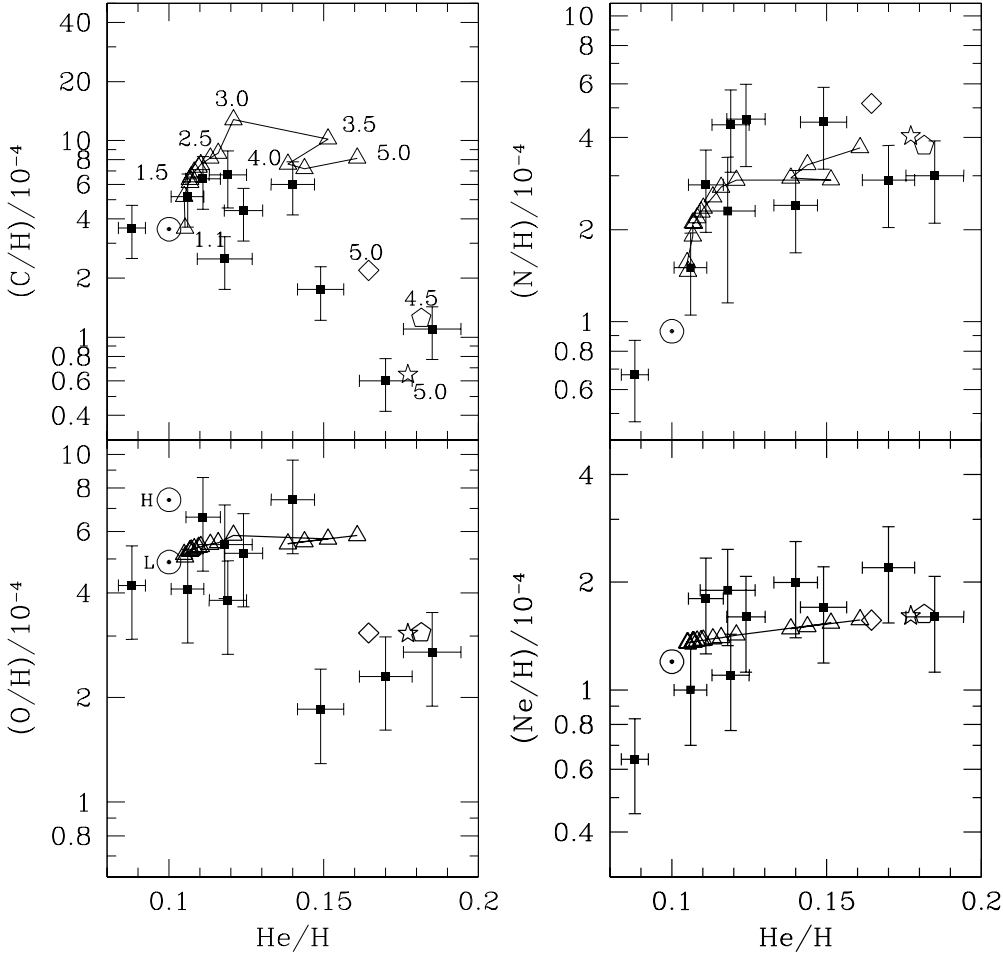}
\end{minipage}
\hfill
\begin{minipage}{0.21\textwidth}
\caption{Summary of the results that best fit the 
observed PN abundances. 
Specifically, triangles represent the same  predictions
as in Fig.~\protect{\ref{fig1_z02lowmkvar}}, that are 
derived from  TP-AGB models with $Z=0.019$ and 
with initial stellar masses from 1.1 to 5.0 $M_{\odot}$
The other three symbols correspond to  intermediate-mass
stars with initial LMC composition ($Z=0.008$), namely:
pentagon for the $4.5 M_{\odot}$ model with K) prescriptions,
star for the  $5.0 M_{\odot}$ model with K) prescriptions, and
diamond for the  $5.0 M_{\odot}$ model with J) prescriptions.
See text for more details}
\label{fig1_z02kvar}
\end{minipage}
\end{figure*}
 
\begin{figure*}
\includegraphics[width=\textwidth]{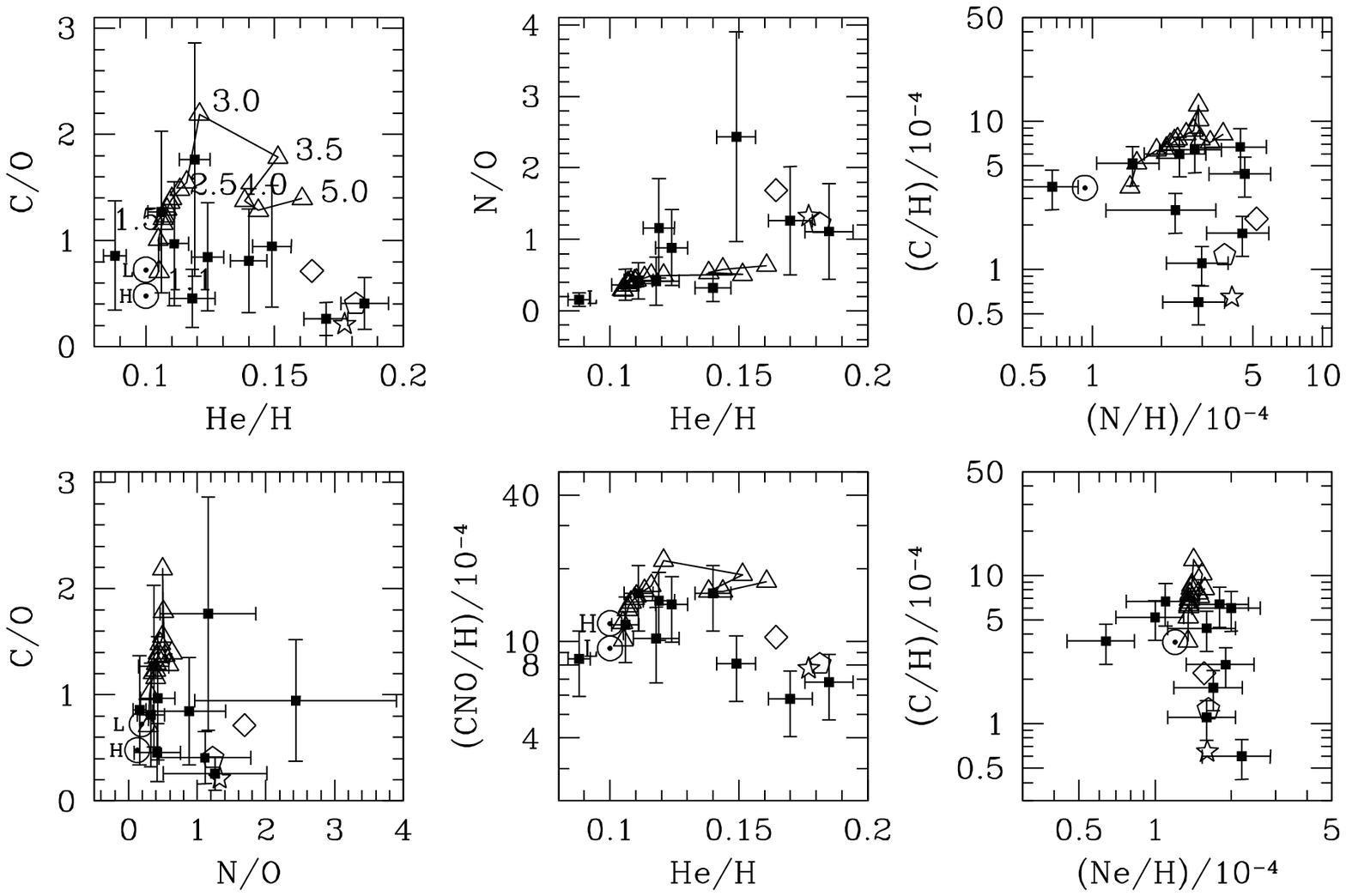}
\caption{The same as in Fig.~\ref{fig1_z02kvar},
but expressing the abundance data with different combinations 
of elemental ratios}
\label{fig2_z02kvar}
\end{figure*}

 
By looking at Fig.~\ref{pn1_z02_d12},
we  conclude  that model 
predictions after the first and second dredge-up processes 
cannot account for the extremely high He/H values of these PNe.
Then, we are led to consider  additional He contributions
from the third dredge-up and HBB during the TP-AGB phase.

Qualitatively, the enrichment in helium and nitrogen and
the simultaneous deficiency in carbon would naturally point to HBB as a 
possible responsible process, via the CNO-cycle reactions.
Therefore, as a working hypothesis, let us assume that 
the stellar progenitors of
these extremely He-rich PNe are intermediate-mass stars 
($M \ga 4.5 \, M_{\odot}$), experiencing 
HBB during their AGB evolution.
We will now  investigate under which conditions all elemental 
features can be reproduced.

To allow an easier understanding of the following analysis, 
Figs.~\ref{fig_hbbz02} and \ref{fig_hbbz008} show
the predicted time evolution 
of He, C, N, O, and Ne elemental abundances in the envelope during
the TP-AGB phase of models experiencing HBB.
Different assumptions are explored. The observed PN abundances
should be compared with the last starred point along 
the theoretical curves, which
marks the last event of mass ejection, and it  
may be then considered representative of the expected PN abundances.    
 
\subsubsection{Constraints from oxygen and sulfur abundances} 
\label{sssec_os} 
At this point additional information comes from the marked oxygen
under-abundance compared  to solar 
(for both High and Low values), common to the extremely He-rich PNe.

We recall that the oxygen abundance in the envelope remains 
essentially unchanged after the first and second
dredge-up events. The third dredge-up may potentially increase
oxygen, depending on the chemical composition of the 
convective inter-shell that forms at thermal pulses.
In any case, no oxygen depletion is expected by any of these
processes.  
A destruction of oxygen could be caused  by a very efficient
HBB, that is if the ON cycle is activated and oxygen starts being
transformed into nitrogen.
  
We have explored this possibility on a $5 M_{\odot}$ TP-AGB model
with original solar metallicity. 
To analyse the effects of a larger HBB efficiency, the mixing-length
parameter $\alpha_{\rm ML}$ has been increased,
and set equal to 1.68, 2.00, and 2.50.
In fact, larger values of $\alpha_{\rm ML}$ correspond to
higher temperatures at the base of the convective envelope.
In none of the three cases have we found any hint of oxygen destruction,
as indicated by the flat behaviour of the abundance curves 
in the bottom-left panel of Fig.~\ref{fig_hbbz02} 
(solid  and long-dashed lines, for $\alpha=1.68$ and 2.50, 
respectively). 

At this point we decided to stop further increasing $\alpha$
-- which would have likely led to oxygen destruction at some point -- 
since we run into a major discrepancy.
In fact, increasing the efficiency of HBB causes 
 a systematic over-enrichment in nitrogen, as shown by the model 
with  $\alpha_{\rm ML}=2.50$ (short-dashed
line). We also note that 
a significant nitrogen production is accompanied by a mirror-like 
destruction of carbon (upper-left panel of Fig.~\ref{fig_hbbz02}).
This is not the case of the model with  $\alpha_{\rm ML}=1.68$
(solid line), in which HBB is almost inoperative.

From these results we can expect that, even if 
a destruction of oxygen is obtained for larger values
of $\alpha_{\rm ML}$, the problem of nitrogen over-production 
would become even more severe.
All these considerations suggest that 
the most He-rich PNe in the sample
should not descend from stars with original 
solar metallicity, but rather from more metal-poor progenitors.
In other words, the observed sub-solar oxygen abundances likely reflect
the initial stellar metallicity.

To test the hypothesis of an original lower
oxygen content we perform explorative TP-AGB evolutionary calculations 
of intermediate-mass stars with initial LMC composition, 
characterised  by roughly half-solar metallicity.
The predicted envelope abundances after the first and second dredge-up 
for stellar models with $[Z=0.008, Y=0.25]$ 
are shown in Fig.~\ref{pn1_z02_d12}. 
As we see the oxygen curve (bottom-left panel) 
is now lower than the corresponding one
for solar composition, and it appears to be much more 
consistent, but not fully, with the location of the He-rich PNe. 
In the next Sects.~\ref{sssec_cn} and \ref{sssec_ne}, we will
test whether these models 
are actually able to fulfil, besides the oxygen data, 
all other chemical constraints 
related to He, C, N, and Ne abundances.
Finally, we note that even better results for O may be obtained 
by adopting an initial oxygen abundance for the LMC composition, 
scaled from the new solar determination by Allende Prieto et al. (2001).

It is worth adding now some  considerations about the 
possibility that the most He-rich PNe in the sample evolved from 
stars with sub-solar metallicity.

  On the one side, our proposed scenario may run into trouble if we
  suppose that the history of chemical enrichment of our Galactic disk
  simply follows an age-metallicity relation, in which younger ages
  correspond to larger metallicities.  In fact in this case, we would
  expect that stars with initial masses as large as $\sim 5\,
  M_{\odot}$ should form from gas with comparable, or even higher,
  degree of metal enrichment than the Sun.
  However, more detailed considerations show that
  the assumption of a unique age-metallicity relation provides an
  over-simplified description
  of the actual chemical evolution in the Galactic disk.
  
  For instance, the Orion nebula has a lower metallicity than solar
  even though it is younger. Also, in the solar neighborhood
  differences do exist. Edvardsson et al. (1993) studied a sample of nearly 
  200 F and G dwarfs in the Galactic disk and found a considerable [Fe/H]
  scatter even for stars with similar age and belonging to a nearby field
  of the disk.  They concluded from this that the chemical enrichment in
  the Galaxy is inhomogeneous. There is therefore a fraction of disk stars in 
  the sky with different composition from solar.  The three PNe with 
  high He abundance belong to the disk and are indeed
  in the same part of the sky (within $\pm$10\degr~of the
  Galactic center). We may reasonably state that these objects belong to
  lowest tail of the metallicity distribution in the Galactic disk.
  These arguments support the use of a metallicity different from
  solar but do not point to LMC metallicity. Another
  weak point is that the metallicity generally increases towards the
  center of the Galaxy, while
  these PNe should have a lower metallicity.
  
  An additional hint for a different initial composition from solar is
  given by sulfur.  All PNe in the sample display a clear
  under-abundance of sulfur with respect to solar (see
  Fig.~\ref{solar}), whereas this element is expected not to change in
  the course of stellar evolution. Its abundance in our PN sample is
  even lower than in Orion (see Table\,\ref{abun}), which is a young
  stellar object.  This implies that either the nucleosynthesis
  history of sulfur is not yet understood and it can actually be destroyed in
  the course of the evolution, or that the initial composition, at
  least of this element, can deviate from solar. Alternatively, the
  determination of the solar abundance may be wrong.  This
    possibility seems the most plausible. The sulfur
  abundance is within the range of values given by
  Mart\'{\i}n-Hern\'andez et al. (2002) for a sample of galactic H\,II
  regions. If the initial abundance of sulfur is not solar, this may
  also be the case for the other elements. Maybe the He-rich PNe are 
  consistent with sub-solar metallicities but, since all observed PNe are 
  under-abundant in sulfur, we are left to explain why
  for some PNe we need solar metallicity whereas for
  others a lower metallicity is invoked,
  this latter feature
  only applying to the most He-rich objects.
  This might be the result of some, still unidentified, selection effect.
   
  The PNe with high helium abundance also show the lowest
  (C+N+O)/H abundance (see Fig.\,\ref{solar}). This, again,  
  suggests a lower initial metallicity. Nevertheless the neon, sulfur
  and argon abundances are similar to the rest of planetaries 
   which would in principle be against this suggestion. The similarity in
   neon of these three PNe with the rest can be explained invoking the
  production  of neon in the course of evolution  (see
  Sec.\,6.4.3). The LMC sulfur abundance is  compatible
  with all the PNe in the sample. As discussed above it might be that
  the sulfur solar abundance is wrong since this is a primary element
  and the observed abundances are quite accurate. The
  argon is more tricky. All PNe show similar argon abundances which are actually
  close to solar. The LMC argon abundance is much lower than  that
  of the high helium PNe. The discrepancy between the subsolar sulfur
  and the solar argon abundance is not readily explained.

  These aspects still need to be clarified, but we would
  remark that a possible answer may come in the
  context of a dedicated study of the chemical evolution
  of our Galaxy, which is beyond the scope of the present work.

\subsubsection{Constraints from carbon and nitrogen abundances}
\label{sssec_cn} 
Perusal  of the C/H and N/H curves in Fig.~\ref{pn1_z02_d12}
(predicted after the first  and second dredge-up) for the LMC composition,
shows that they share the trends with the stellar 
mass as the solar-metallicity curves, 
but the carbon and nitrogen abundances are now systematically lower, 
as for oxygen.
In particular  carbon abundances, prior to the TP-AGB evolution,  
-- at any stellar mass -- are now 
compatible with the low values measured in the He-rich PNe.
As for nitrogen,  some enrichment  
seems instead required (up to a factor of 2 at the largest
stellar masses) to obtain agreement with the observations.

Given these premises, a number of TP-AGB models with $Z=0.008$ 
and masses of $4.0 - 5.0 M_{\odot}$ have been calculated 
for different choices of parameters, in view to simultaneously
matching the observed carbon, nitrogen and helium abundances.
We just summarise the main emerging points.

\paragraph{Efficiency of the third dredge-up.}

High He abundances strongly depend on the
efficiency of the third dredge-up.
Figure~\ref{fig_hbbz008} shows that 
relatively low values of $\lambda$ lead  
to inadequate helium enrichment, because both a limited amount of 
helium is dredged-up at each thermal pulse, and the duration 
of the TP-AGB phase becomes shorter, implying fewer dredge-up events.
This is true for the TP-AGB models with 
$M = 4.5,5.0\, M_{\odot}$, and $\lambda=0.5$ 
(dotted and dot-short-dashed curves), which do not go beyond He/H$=0.13$. 

On the other hand, TP-AGB models of the same initial masses 
but with $\lambda \sim 0.9$ become highly enriched in helium, 
yielding a very close agreement 
with the observed data for He/H (solid, short-dashed, and long-dashed curves). 
The largest He/H values measured in PNe 
are reproduced under the assumption 
that  massive TP-AGB stars experience, besides HBB, 
a large number (of the order of $100$) 
of very efficient third dredge-up episodes.

This indication  is supported by recent  results 
of full TP-AGB calculations for stellar masses $\ga 5 M_{\odot}$
(Vassiliadis \& Wood 1993; Frost et al. 1998; Karakas et al. 2002).

\paragraph{The problem of nitrogen over-production.}

Invoking a large dredge-up efficiency does 
not guarantee, however, that all other elemental features are
reproduced.  
In fact, assuming  a very efficient dredge-up $\lambda \sim 0.9$,  
{\em and} the standard intershell chemical
composition (i.e. $X_{\rm csh}(^{12}{\rm C})= 0.22$, 
$X_{\rm csh}(^{16}{\rm O})= 0.02$, $X_{\rm csh}(^{4}{\rm He}) =0.76$; 
see Sect.~\ref{ssec_modpre}), we find 
a sizeable over-production of nitrogen by HBB.
This is illustrated in Fig.~\ref{fig_hbbz008} by the N/H curve 
corresponding to the $5 M_{\odot}$ model 
(with $\kappa_{\rm fix}$ prescription; dot-long-dashed line).
Calculations were stopped before the termination 
of the TP-AGB evolution (i.e. before the ejection of the entire envelope),
since the He/H in the envelope already exceeded 0.2 and 
the over-production of nitrogen was very large.     

In relation to this latter point, going back to the past literature, 
the same problem was already pointed out by Marigo et al. (1998; see their 
section 5.4) and earlier by Becker \& Iben
(1980; see their section VII), 
in their careful analysis on the expected abundance
variations during the TP-AGB phase of intermediate-mass stars and the
observed PNe abundances (see their section VII).  These authors
demonstrated that the efficiency of HBB, required to achieve the small
C/O values exhibited by PNe with large He/H, is such as to yield N/O
ratios that exceed by over an order of magnitude the PN values.
This indication is quite robust 
and almost model-independent, since
it merely reflects the interplay  between nuclear reactions 
of the CN-cycle.   
The efficiency of the third
dredge-up directly affects the HBB nucleosynthesis: if the CN-cycle
operates at equilibrium (a condition often met by stars with HBB)
the more carbon that is dredged-up, the more nitrogen is eventually
synthesised.

After discussing various aspects of the issue, Becker \& Iben (1980)
concluded that the positive correlation  
between N/O and He/H observed in helium-rich PNe could be reproduced 
``by supposing that $^{12}{\rm C}$ is burned at a modest rate in the 
convective envelope'' of the most massive TP-AGB stars.
In other words, they invoked a weak efficiency of HBB.
However, these authors also clearly stated that in this way one would 
at the same time cope  with the discrepancy of predicting too large atomic C/O 
ratios, contrary to observations.  
The only possible explanation, plausible at that time, to reconcile 
all points was the hypothesis of a significant dust-depletion of carbon,  
with consequent apparent decrease of the measured C/O ratio
(involving the atomic carbon).
In this study we explore other possibilities, as reported below.

\paragraph{The effect of variable molecular opacities.}
 
First we consider the effect of variable molecular opacities.
Replacing the $\kappa_{\rm fix}$ with $\kappa_{\rm var}$ prescription
in the $5 M_{\odot}$ model (while keeping $\lambda=0.9$, 
and the standard chemical composition), the evolution of He, C, N 
surface abundances change dramatically. In practice, we pass from  
the problem of a huge nitrogen over-production 
for the $\kappa_{\rm fix}$ case (dot-long-dashed N/H curve)
to those of almost zero nitrogen synthesis and 
carbon over-enrichment for the $\kappa_{\rm var}$ model 
(short-long-dashed line).

The latter result is due to the weakening, 
or even prevention, of HBB in intermediate-mass stars
that undergo efficient carbon enrichment by the third dredge-up
during the early stages of their  TP-AGB evolution. As discussed 
by Marigo (2003), the increase in molecular opacities,
as soon as C/O becomes larger than one, causes cooling 
at the base of the convective envelope, which
may extinguish the CNO-cycle reactions associated with HBB.

Both  models with different opacities do not 
reproduce the observed PN data for C and N, though  both 
seem to imply the same direction: too much carbon is assumed to be 
injected  by the third dredge-up into the convective envelope. 
This causes the over-production of nitrogen in the $\kappa_{\rm fix}$ model,
while it yields a net over-enrichment of carbon 
in the $\kappa_{\rm var}$ model.

These considerations introduce us to another possibility to solve
this  problem, related to the chemical composition
of the convective inter-shell. 
    
\paragraph{Inter-shell chemical composition.}

We propose here a solution, different from that 
suggested by Becker \& Iben (1980),
to account simultaneously for the observed  N/O and He/H correlation
\underline{and}  C/O and He/H anti-correlation.
It is based on the assumed elemental abundances -- essentially  $^{4}{\rm He}$,
$^{12}{\rm C}$, and $^{16}{\rm O}$ --
in  the convective inter-shell developed at thermal pulses, part of which
is then dredged-up to the surface.

We find that all discrepancies -- relative to the N and/or C over-production --
are removed by  relaxing the usual prescription 
of the standard inter-shell chemical composition (Boothroyd \& Sackmann 1998, 
see also Sect.~\ref{ssec_modpre},  and Table~\ref{tab_mod}), and 
assuming instead that the dredged-up material in intermediate-mass TP-AGB
stars consists mainly of helium, with very little carbon and practically no oxygen.
Results are shown in Fig.~\ref{fig_hbbz008} (solid, short- and long-dashed
curves).
Thanks to the small amount of dredged-up carbon, 
we would favour the enrichment in helium and avoid 
the nitrogen over-production.

These indications have been initially derived 
from empirical evidence, and it is encouraging to receive  
also some theoretical support from    
calculations of the TP-AGB evolution of intermediate-mass stars.
To our knowledge, it was  first mentioned  -- that the typical chemical 
inter-shell composition may be quite different from the standard one 
in the most massive TP-AGB stars -- in the work of 
Vassiliadis \& Wood (1993). The authors point out that 
the thermal pulses in TP-AGB stars as massive 
as $5\, M_{\odot}$ 
are characterised 
by extremely deep dredge-up ($\lambda \approx 0.8$), 
and weak efficiency of the 
triple-$\alpha$ reaction in the convective inter-shell due to both 
the rapid quenching of the instability.
These authors state that
in their calculations  the  dredged-up material mainly 
consists of helium and nitrogen
produced by the CNO cycle.
Consequently, this result should imply a lower synthesis of primary carbon
in the convective inter-shell, hence a lower carbon abundance in 
the dredged-up material compared to the standard values 
($X_{\rm csh}(^{12}{\rm C}) <  0.2-0.3$).  This indication agrees 
with  our suggestion to explain
the measured CNO abundances in PNe with very high helium content.

 A few years later, Frost et al. (1998) 
partly confirmed the results by  Vassiliadis \& Wood (1993),
and pushed the analysis further reporting the discovery 
of a new kind 
of thermal pulses,
designated as ``degenerate thermal pulses'' 
(the reader should refer to that work for all details). 
In few words a massive 
TP-AGB star would  experience a sort of cyclic trend: 
First a relatively large number of weak thermal 
pulses (say $30-40$) with very deep dredge-up takes place 
leaving a long tail of unburned helium, 
that is then burned at once by  
a strong ``degenerate pulse'' 
(the carbon abundance in the inter-shell reaches $^{12}{\rm C} \approx  0.6$), 
after which another  sequence of 
weak pulses starts again, and so on.
Similar trends have been reported by Siess et al. 
(2002) in their study of the TP-AGB evolution in population III stars.

It should be remarked that while there is a close agreement in the
results by Frost et al. (1998) and Vassiliadis
\& Wood (1993) with respect to the high efficiency of the third dredge-up
($\lambda \approx 1$) in these massive AGB stars, some substantial differences
could be present instead in the predicted chemical composition 
of the dredged-up 
material.
In fact, the possibility of a carbon abundance lower than
the standard value ($X_{\rm csh}(^{12}{\rm C}) < 0.2-0.3$), which is deduced 
from  the Vassiliadis \& Wood's paper,  
is actually not confirmed by the work of Frost et al. 
(1998; also Lattanzio's private communication).
These possible differences 
could be ascribed to different physical and numerical
details of the stellar evolution codes, that can significantly
affect the results (see Frost et al. 1998; Frost \& Lattanzio 1996; and 
Lattanzio's private communication).  

We believe that this point is crucial and it deserves a clarification 
with the aid of full calculations of the thermal pulses experienced
by the most massive AGB models.
Anyway, as our study represents a sort of empirical 
calibration of the AGB nucleosynthesis,  we follow Vassiliadis \& Wood's 
indications and assume 
the possibility that during their TP-AGB
evolution the most massive  intermediate-mass stars suffer many deep 
dredge-up events that bring a large amount 
of helium, but little carbon up to the surface. 
In fact, as mentioned above, this seems just 
what we need to reproduce the observed abundances of the most He-rich PNe, 
thus solving a long-standing problem already stated by Becker \& Iben (1980).

The quantitative confirmation comes from the results of our TP-AGB calculations
presented in  Fig.~\ref{fig_hbbz008}, referring to models with i) 
stellar masses in the range $4.5-5.0\, M_{\odot}$, ii) initial LMC composition ($Z=0.008$),
iii) assumed very deep dredge-up ($\lambda \sim 0.9$), and iv) inter-shell chemical
composition with typically ($X_{\rm csh}(^{12}{\rm C})= 0.02-0.03$, 
$X_{\rm csh}(^{16}{\rm O})= 0.002-0.003$, $X_{\rm csh}(^{4}{\rm He})=0.967-0.978$).
These models experience also HBB (more efficient at larger masses)
as we see by considering the mirror-like evolution of the C- and N-curves in the
top-left and top-right panels, respectively. The final points, which would 
correspond to the predicted PN abundances, are clearly in  very good agreement
with the observed data for the most He-rich PNe. In particular, models are
able to reach He/H$\sim 0.17-0.20$, without over-producing nitrogen, and accounting
for the extent of carbon depletion. We note that the 
rising part of the C-curves towards the end of the evolution, and the concomitant
flattening of the N-curves reflect the eventual extinction of HBB while the last 
dredge-up events still take place.

\subsubsection{Constraints from neon abundances}
\label{sssec_ne} 
Within the lowest extension of the error-bars,
the data for the most He-rich PNe do not show a significant overabundance 
with respect to solar, though some increasing trend of Ne/H with
He/H might be present.
Some useful indications 
can be derived.
 
Recalling that both the first and second dredge-up episodes are expected
not to alter the initial neon abundance, and   
looking at Fig.~\ref{pn1_z02_d12} we point out two facts, namely:
i) the predicted Ne/H after the second dredge-up in the most massive
stars (with $M \sim 4-5\,  M_{\odot}$) with initial solar composition are
already compatible with the measured Ne/H in the most He-rich PNe, whereas 
the models  of the same stellar mass but with LMC composition lie clearly
below the observed points;
ii) the slightly increasing trend of  Ne/H with He/H at increasing stellar mass
merely reflect the decrease in the surface H content due to the dredge-up
events, since the neon abundance is completely unaffected.

The consideration of these two points, together with the conclusions from the 
analysis carried out in the preceding sections, would point to 
a significant Ne production having occurred in the progenitor
stars of the most He-rich PNe, since
all other PN abundances (He, C, N, and O) seem
globally consistent with stars originating from gas of sub-solar metallicity.
As shown, our intermediate-mass models with LMC composition do  satisfy all 
these chemical constraints.

Of course, another possibility would be 
$({\rm Ne}/Z)_{\rm LMC} > ({\rm Ne}/Z)_{\odot}$, that is 
the initial content of Ne in the original metal-poor gas was enhanced 
 compared to what is expected for a solar-metallicity-scaled mixture. 
In this case, there is no  necessity that Ne 
is synthesised during the AGB phase.

Going back to the former alternative, which invokes a sizeable
Ne production in intermediate-mass stars with LMC composition, we find that
models are able to attain a good agreement with the Ne/H values measured in
the most He-rich PNe by assuming that 
the enrichment of neon is due to the synthesis of $^{22}$Ne
during thermal pulses via the chain of reactions
$^{14}$N($\alpha$, $\gamma$)\, $^{18}$F($\beta^+$, $\nu$)\,
$^{18}$O($\alpha$, $\gamma$)\, $^{22}$Ne. 
In other words, all nitrogen in the inter-shell should be converted 
into $^{22}$Ne {\bf (see Sect.~\ref{ssec_modpre}).}
The possible channel of subsequent destruction
through  $^{22}$Ne($\alpha$, n)$^{25}$Mg should be inefficient.
In our calculations we assume that just $1 \%$ of the newly synthesised 
$^{22}$Ne is burned into $^{25}$Mg (see also Sect.~\ref{ssec_modpre}).

Then, in the context of the proposed interpretation, the immediate 
consequence would be the inefficiency 
of the $^{22}$Ne($\alpha$, n)$^{25}$Mg reaction
as production channel of neutrons, with important implications
for the synthesis of slow-neutron capture elements (Busso et al. 1999). 
In this case, the major role for the s-process nucleosynthesis would be played 
by the $^{13}$C($\alpha$, n)$^{16}$O channel.

\section{Summary and conclusions}
\label{sec_concl}
In this study a sample of PNe  with accurately determined elemental
abundances, is used to derive interesting information about the evolution
of the stellar progenitors, and set constraints on the nucleosynthesis
and mixing processes characterising their  previous evolution.
To this aim, synthetic TP-AGB models are calculated to
reproduce the data by varying the parameters:
initial stellar mass and metallicity, molecular opacities,
dredge-up and HBB efficiency, and chemical composition of the convective 
inter-shell developed at thermal pulses. 

The  clear segregation of the abundance data in two sub-samples,
particularly evident in the O/H -- He/H diagram,  
has led us to discuss them separately. And indeed, our investigation
suggests two different interpretative scenarios.
The final results that best reproduce the observed data are  
summarised in Figs.~\ref{fig1_z02kvar} and 
\ref{fig2_z02kvar}. 
   
From the analysis of the
group of PNe with low He content  (He/H$<0.15$) 
and solar-like oxygen abundances, we conclude that:
\begin{itemize}
\item The stellar progenitors are low- and intermediate-mass 
stars with original solar-like chemical composition and initial masses 
spanning the range $0.9-4.0\, M_{\odot}$. 
\item The oxygen abundances are consistent 
with the recent determination for the Sun by Allende Prieto et al. (2001), 
that is lower than previous estimates (see e.g. Anders \& Grevesse 1989) 
by almost 0.2 dex.
For a few PNe there may be a limited oxygen enrichment,
possibly associated with dredge-up during the TP-AGB phase. 
\item There is clear evidence of carbon enrichment in some PNe that also
exhibit C/O$>1$, suggesting that they evolved from carbon stars experiencing 
the third dredge-up during the TP-AGB phase. 
\item Measured carbon abundances are well reproduced by TP-AGB models 
with dredge-up efficiencies $\lambda \sim 0.3-0.4$, 
and adopting variable molecular opacities in place of the usual
solar-scaled opacity tables (see Marigo 2002).
The introduction of variable 
opacities prevents the likely over-enrichment of carbon 
by shortening the duration of the carbon-star phase, and causing an earlier
shut-down of the third dredge-up due to the cooling of the envelope structure 
(Marigo 2003).

\item The degree of nitrogen enrichment is consistent with the expectations 
from
the first and second dredge-up events, occurred prior to the TP-AGB
phase. The efficiency of HBB in intermediate-mass stars with 
solar-metallicity should be modest.
\item Helium abundances are well accounted for by considering the 
whole contribution of all dredge-up processes (i.e. first, and possibly second 
and third).
\end{itemize}          

From the study of the extremely helium-rich  ($0.15\le $He/H$\le 0.20$) and 
oxygen-poor PNe we can conclude the following:
\begin{itemize}
\item The stellar progenitors should be intermediate-mass stars
($4-5\,M_{\odot}$) experiencing both the third dredge-up and HBB during
their TP-AGB evolution.
\item The PN oxygen abundances are consistent with 
a sub-solar initial stellar metallicity. 
In fact, under the hypothesis of solar metallicity we are forced to 
invoke a significant oxygen destruction via very efficient HBB, which  
violates other chemical constraints, e.g. causing a large over-production
of nitrogen. 
Instead, models with assumed initial LMC composition provide a fairly 
good agreement with the data. 

\item The first two assumptions are needed in the models to
    reproduce the observed abundances. This leads to a controversy
    which is important to point out. The combination of low
    metallicity with intermediate-mass progenitors is peculiar, since
    these stars are probably recently formed from gas with
    interstellar abundances.  Although there are several indications
    that these PNe are of lower metallicity (see Sec.\,6.4.1), this
    could also suggest that perhaps another physical process has not
    fully been taken into account.  This issue should be further
    investigated.
 
\item The long-standing problem -- initially formulated by 
Becker \& Iben (1980) -- of accounting, simultaneously  and quantitatively,
for the observed N/O--He/H correlation and the C/O--N/O anti-correlation
seems to be solved by assuming that the third dredge-up i) 
is very efficient, and 
ii) brings up to the surface material containing only a small amount of 
primary carbon synthesised during thermal pulses. A good agreement with 
the observed data is obtained by adopting $\lambda \sim 0.9$ and 
$X_{\rm csh}(^{12}{\rm C})\sim 0.02-0.03$.

The former indication on the dredge-up efficiency, 
derived empirically, 
is supported  on theoretical grounds 
by full TP-AGB calculations of intermediate-mass stars, i.e.
Vassiliadis \& Wood (1993), and more recently  Frost et al. (1998) 
and Siess et al. (2002). A significant part of the  
helium enrichment of these PNe  should 
be ascribed to a large number of deep 
dredge-up events that precede the occurrence of 
the so-called 
``degenerate pulses'', according to the designation 
introduced by Frost et al. (1998). 
The additional requirement emerging from our study --  
that such dredge-up events should not only be extremely 
deep but also carry a small amount of carbon -- is not fully confirmed 
by theoretical analyses (i.e. Frost et al. 1998), 
though a positive indication in this sense
is given by the work of Vassiliadis \& Wood (1993).

\item A significant production of $^{22}$Ne -- via $\alpha$-captures starting 
from $^{14}$N -- should take place in these stars to reproduce the observed 
Ne/H values of the He-rich PNe, under the hypothesis they descend from 
intermediate-mass stars with initial LMC chemical composition.
As direct consequence, this would imply a reduced role of the 
$^{22}$Ne($\alpha$, n)$^{25}$Mg reaction in providing neutrons for the
slow-neutron capture nucleosyntesis that is expected to occur during thermal
pulses.

We also note that the invoked inefficiency 
of the $^{22}$Ne($\alpha$, n)$^{25}$Mg 
channel seems consistent with the expected low synthesis 
of carbon at thermal pulses in the most massive AGB stars 
experiencing very deep dredge-up (see former point). 
In fact, as reported by 
Vassiliadis \& Wood (1993) 
the deep dredge-up quickly extinguishes the 
helium burning shell. As a consequence this could 
prevent both a significant production of 
carbon via the triple-$\alpha$ reaction and 
the attainment of the high temperatures required
for the full activation of the $^{22}$Ne($\alpha$, n)$^{25}$Mg reaction.
The confirmation of this two-fold aspect deserves 
detailed  calculations of thermal pulses. 
\end{itemize}

\begin{acknowledgements}
P.M. acknowledges the SRON National Institute for Space Research (Groningen) 
for hospitality and financial support during her visit in October 2001,  
and the Italian Ministry of Education, University and Research (MIUR)
for the work carried out at the Astronomy Department in Padova.
J. Bernard-Salas thanks Annette Ferguson for valuable discussions and
suggestions as well as the Universit\`a di Padova for hospitality. We
are grateful to Dr. John Lattanzio for the careful 
reading of this manuscript, and for providing comments and suggestions  
that result in an improvement
of the paper.

\end{acknowledgements}

\end{document}